\documentclass[preprintnumbers,
twocolumn,
prd,
aps,superscriptaddress,footinbib,amsfonts,amsmath,amssymb,showpacs,floatfix,nofootinbib]{revtex4-2}

\usepackage{amsmath,amssymb,amsfonts,mathrsfs,bbold}
\usepackage{yfonts}
\usepackage{graphicx}
\usepackage[colorlinks=true,citecolor=blue,linkcolor=blue,urlcolor=blue]{hyperref}
\usepackage[dvipsnames]{xcolor}
\usepackage{lipsum}
\usepackage{slashed}
\usepackage[normalem]{ulem}
\usepackage{url}

\usepackage{amsthm}
\usepackage{physics}
\usepackage{bm}
\usepackage{upgreek}

\newcommand{\ben}{\begin{eqnarray*}}
\newcommand{\een}{\end{eqnarray*}}

\def\vh{\hat{v}}
\def\wh{\hat{w}}
\def\sh{\hat{s}}

\begin{document}
%%%%%%%%%%%%%%%%%%%%%%%%%%%%%%%%%%%%%%%%%%%%%%%%%%%%%%%%%%%%%%%%%

\preprint{
	{\vbox {			
		\hbox{\bf JLAB-THY-25-4341}
}}}
\vspace*{0.2cm}

%================================================================
\title{Factorized QED and QCD Contribution to Deeply Inelastic Scattering}
%================================================================

%--------
\author{Justin Cammarota}
\email{jcammarota@wm.edu}
\affiliation{Department of Physics, William \& Mary,
Williamsburg, Virginia 23187, USA}

\author{Jian-Wei Qiu}
\email{jqiu@jlab.org}
\affiliation{Department of Physics, William \& Mary,
Williamsburg, Virginia 23187, USA}
\affiliation{Theory Center, Jefferson Lab,
Newport News, Virginia 23606, USA}

\author{Kazuhiro Watanabe}
\email{kazuhiro.watanabe.b8@tohoku.ac.jp}
\affiliation{Faculty of Science and Technology, 
Seikei University, Musashino, 
Tokyo 180-8633, Japan}
\affiliation{Department of Physics,
Tohoku University, Sendai 980-8578, Japan}

\author{Jia-Yue Zhang}
\email{jzhang@jlab.org (corresponding author)}
\affiliation{Theory Center, Jefferson Lab,
Newport News, Virginia 23606, USA}

\date{\today}
%================================================================

%================================================================
\begin{abstract}
We present the first calculation of next-to-leading order (NLO) factorized QED and QCD contributions to the short-distance hard coefficients of inclusive lepton-hadron deep inelastic scattering (DIS) in a joint QED and QCD factorization approach.  Unlike the traditional radiative correction approach to handle the collision-induced QED contributions to DIS, QED radiation from all charged leptons and quarks are treated equally, and their collinear sensitivities are systematically factorized into corresponding universal lepton and parton distribution functions.  We demonstrate that the NLO factorized QED contribution is completely infrared safe and calculable without the need of any parameters other than the standard factorization scale in the same way as the factorized QCD contribution.  We discuss
the potential impact of this joint factorization approach on the extraction of partonic information from lepton-hadron DIS.
\end{abstract}
%================================================================

\maketitle

%%%%%%%%%%%%%%%%%%%%%%%%%%%%%%%%%%%%%%%%%%%%%%%%%%%%%%%%%%%%%%%%%

%================================================================
\section{Introduction}
\label{sec:intro}
%================================================================

The experiment of lepton-hadron deep-inelastic scattering (DIS) at Stanford Linear Accelerator Center (SLAC) over 50 years ago~\cite{Bloom:1969kc} discovered the partonic structure of the proton that led to the discovery of the fundamental theory of strong interactions known as Quantum Chromodynamics (QCD) (for reviews, see~\cite{Brambilla:2014jmp,Gross:2022hyw}). Since then, many DIS experiments at lepton-hadron facilities around the world have been carried out and have played a critical role in the development of our understanding of QCD and the strong nuclear force, as well as the internal structure of nucleons and nuclei. With the high luminosity lepton-hadron fixed target facility at the Jefferson Lab, the future electron-ion collider (EIC) to be constructed at the Brookhaven National Lab, and similar facilities around the world, the lepton-hadron scattering with a large momentum transfer will continue to play an important and unique role in our efforts to explore the internal structure of hadrons and to study the emergence of hadrons from produced quarks and gluons in the high energy collisions~\cite{Accardi:2012qut}.

For the inclusive DIS between a lepton of momentum $\ell$ and a hadron of momentum $P$, $e(\ell)+h(P) \to e(\ell') + X$, we measure the scattered lepton of momentum $\ell'$, from which we determine the transfer of momentum $q^\mu=(\ell - \ell')^\mu$. When $Q^2=-q^2$ is sufficiently large ($\gg \Lambda_{\rm QCD}^2)$, QCD factorization~\cite{Collins:1989gx} has been successfully applied to lepton-hadron scattering cross sections by separating perturbatively calculable short-distance physics at ${\cal O}(1/Q)$ from the non-perturbative physics that we want to learn at ${\cal O}(R\sim 1/\Lambda_{\rm QCD})$ with nucleon radius $R$. By varying $Q^2$ and the Bjorken variable $x_B=Q^2/2P\cdot q$, the exchanged gauge boson of momentum $q^\mu$ between the colliding lepton and hadron has been considered as a good and controllable hard probe to explore the QCD dynamics and partonic structure inside the colliding hadron~\cite{Roberts:1990ww,Tung:2001cv}.  

With the large momentum transfer $Q^2$ between the colliding lepton and the hadron, the hard scattering necessarily generates both collision-induced QCD and QED radiation, affecting the measured cross sections.  The QCD factorization systematically takes care of the impact from collision-induced QCD radiation by including the collinear sensitive, non-perturbative and {\it process-independent} contributions into the universal parton distribution functions (PDFs) or fragmentation functions (FFs) if specific hadron(s) are measured in the final state, while showing that all {\it process-dependent} contributions are either calculable in QCD perturbation theory as contributions from short-distance partonic scattering to the measured cross sections or suppressed by the inverse power of the hard scale $Q^2$ that can be neglected when $Q^2\gg \Lambda_{\rm QCD}^2$.
On the other hand, historically, the impact of collision-induced QED radiations from the observed leptons of the lepton-hadron DIS have been evaluated perturbatively as radiative corrections (RC) to the lowest-order Born DIS cross section in QED~\cite{Mo:1968cg,Bardin:1989vz,Badelek:1994uq,Kripfganz:1990vm,Spiesberger:1994dm,Blumlein:2002fy}.  Such an approach often needs to introduce unknown parameter(s) that could affect the predictive power of RCs.  

Furthermore, the collision-induced QED contributions to an observed cross section in lepton-hadron collisions can be sensitive to a specific final state that is measured when we go beyond the inclusive DIS for which only the scattered lepton is observed.  For example, when we measure a final-state hadron in addition to the scattered lepton, $e(\ell)+h(P) \to e(\ell') + h'(P')+X$, semi-inclusive DIS (SIDIS) allows us to study the transverse momentum dependent PDFs (or simply, TMDs) from the events when the momentum imbalance between observed final-state lepton of momentum $\ell'$ and hadron of momentum $P'$ is so much smaller than the hard scale of the collisions $Q$.  The angular distribution between the leptonic plane, defined by the momenta of the colliding and scattered lepton ($\ell$ and $\ell'$) and hadronic plane, defined by the momenta of colliding and produced hadron ($P$ and $P'$), is sensitive to the contributions from different TMDs~\cite{Boussarie:2023izj}.  Consequently, corresponding azimuthal angular modulations could provide a powerful tool to isolate and extract contributions from different TMDs.  However, the collision-induced QED radiations can change both the value and direction of momentum transfer $q^\mu$ between the colliding lepton and the hadron, which can alter the relation between the hadron's TMDs and observed angular distributions between the leptonic and hadronic planes to reduce the effectiveness of azimuthal modulations to separate contributions from different TMDs to the measured SIDIS cross sections~\cite{Liu:2020rvc}.  

A new joint QED and QCD factorization approach was recently proposed to treat collision-induced QED and QCD radiation equally~\cite{Liu:2021jfp}. Like QCD factorization for lepton-hadron DIS, the short-distance hard partonic scattering in this joint QED and QCD factorization approach is also infrared safe and calculable in QED and QCD perturbation theory, while collinear sensitive, non-perturbative and process-independent radiations are systematically factorized into universal lepton distribution functions (LDFs) and lepton fragmentation functions (LFFs) and other non-perturbative contributions are power suppressed by the hard scale $Q^2$.  This joint QED and QCD factorization approach improves the predictive power by removing the need of any unknown parameter(s) for including collision-induced QED radiations other than the standard factorization scale~\cite{Cammarota:2024vxj}.

In this paper, we present the first complete calculation of fully factorized next-to-leading order (NLO) QED and QCD contributions to the short-distance hard coefficients of the leading electron-quark subprocess, $e+q\to e + X$, of inclusive lepton-hadron DIS in the joint QED and QCD factorization approach, along with a well-defined procedure to include available higher order QCD contributions into this joint factorization approach.  Unlike the calculation of RCs in the literature, which focuses on photon radiation from the leptons, we evaluate QED radiation from all charged leptons and quarks at the partonic scattering level, and systematically factorize all collinear sensitive radiations from leptons and quarks to corresponding lepton and parton distribution functions, respectively.  We demonstrate that our results are completely infrared safe, only depending on the standard factorization scale, while all leading power non-perturbative and collinear sensitive radiative contributions are included into universal LDFs and LFFs, as well as the PDFs of colliding hadron.  

It is important to stress that collision-induced radiation in this joint QED and QCD factorization approach is nonperturbative since a radiated photon could become a quark-antiquark pair at a nonperturbative scale in addition to becoming a lepton-pair via QED interaction.  We also discuss and explore the uncertainty of LDFs and LFFs and their impact on the precision of theoretical calculations for inclusive lepton-hadron scattering.  

The rest of this paper is organized as follows. In Sec.~\ref{sec:formalism}, we introduce the joint QED and QCD factorization formalism for inclusive DIS.  In Sec.~\ref{sec:qcd}, we present the next-to-leading order (NLO) QCD contribution in this joint QED and QCD factorization approach, including a well-defined procedure to add available QCD calculations beyond the NLO.  In Sec.~\ref{sec:qed}, we carry out calculations of the full NLO QED contributions to the factorized short-distance coefficients of the leading subprocess, $e+q\to e + X$, in this joint QED and QCD factorization approach.  We explore the numerical impact of the NLO QED contributions to the inclusive DIS and the role of universal LDFs and LFFs in Sec.~\ref{sec:numerical}.  Finally, in Sec.~\ref{sec:conclusion}, we provide a summary of the key results of this paper and an outlook of the impact of this joint QED and QCD factorization approach for studying high-energy lepton-hadron collisions.

%================================================================
\section{Joint QED and QCD factorization for lepton-hadron DIS}
\label{sec:formalism}
%================================================================

\begin{figure}[htbp]
	\centering
	 \includegraphics[width=0.16\textwidth]{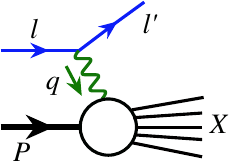} 
	 \hskip 0.25in
	  \includegraphics[width=0.19\textwidth]{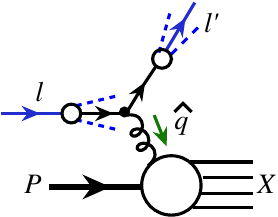} \\
	  (a) \hskip 1.5in (b)
	\caption{(a) Inclusive DIS with one-photon exchange; 
	              (b) Inclusive DIS with collision-induced radiation from colliding and observed leptons.
	}	
\label{fig:dis0}
\end{figure}

The historical lepton-hadron DIS is effectively an inclusive production of a single lepton with its transverse momentum $\ell'_T\propto \sqrt{Q^2}$ much larger than $\Lambda_{\rm QCD}$ in high-energy lepton-hadron collisions.  At the lowest order in QED coupling (the Born cross section), the scattering is approximated by exchanging one virtual photon (or a $Z^0$ boson that is more relevant for DIS with a larger momentum transfer or the parity violating DIS), as shown in Fig.~\ref{fig:dis0}(a).  
%----------------------------------------------------------------
% Figure: Inclusive DIS
%----------------------------------------------------------------
%----------------------------------------------------------------
With this approximation of one-photon exchange, Born cross section in QED can be expressed in terms of a hadronic tensor, $W_h^{\mu\nu}$ and the lowest order leptonic tensor $L_{\mu\nu}^{(0)}$~\cite{Roberts:1990ww,Tung:2001cv}, 
\begin{equation}
E'\frac{d\sigma^{\rm Born}_{e h\to e X}}{d^3\ell'} 
= \frac{2\alpha_{em}^2}{S(Q^2)^2} \,
L^{(0)}_{\mu\nu}(\ell,\ell')\, W_h^{\mu\nu}(q,P)
\label{eq:dis0}
\end{equation}  
where the fine-structure constant $\alpha_{em}=e^2/(4\pi)$, $S=(\ell+P)^2\approx Q^2/(x_B\, y)$ with $y=P\cdot q/P\cdot \ell$. 
In Eq.~(\ref{eq:dis0}), the lowest order unpolarized leptonic tensor is given by 
$L^{(0)}_{\mu\nu}=2(\ell_\mu\ell'_{\nu}+\ell'_\mu\ell_{\nu}-\ell\cdot\ell' g_{\mu\nu})$, 
and corresponding hadronic tensor for an unpolarized hadron, 
\begin{equation}
W_h^{\mu\nu}(q,P) = 
-\widetilde{g}^{\mu\nu} \, F_1(x_B,Q^2)
+\frac{\widetilde{P}^\mu \widetilde{P}^\nu}{P\cdot q}\, F_2(x_B,Q^2)
\label{eq:sfs}
\end{equation}  
with $\widetilde{g}^{\mu\nu} = g^{\mu\nu} - q^\mu q^\nu/q^2$, $\widetilde{P}^\mu = \widetilde{g}^{\mu\nu} P_{\nu}$
and nonperturbative structure functions $F_1$ and $F_2$.  
The unpolarized Born cross section can then be expressed in terms of the structure functions evaluated at $(x_B,Q^2)$,
\begin{eqnarray}
E'\frac{d\sigma^{\rm Born}_{e h\to e X}}{d^3\ell'} 
&=& \frac{4\alpha_{em}^2}{y(Q^2)^2} \bigg[
x_B\,y^2 F_1(x_B,Q^2) 
\label{eq:dis0_sfs} \\
&\ & \hskip 0.2in
+ (1-y-\frac{\gamma^2y^2}{4})F_2(x_B,Q^2)\bigg]
\nonumber
\end{eqnarray}
where $\gamma=2x_BM/Q$ with hadron mass $M$ and $Q=\sqrt{Q^2}$.

\begin{figure}[htbp]
	\centering
	 \includegraphics[width=0.15\textwidth]{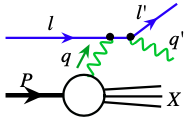} 
	 \hskip 0.4in
	  \includegraphics[width=0.15\textwidth]{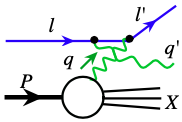} \\
	  \caption{Diagrams contribute to the RC of inclusive DIS cross sections.
	}	
\label{fig:dis1}
\end{figure}

\begin{figure}[htbp]
	\centering
	 \includegraphics[width=0.26\textwidth]{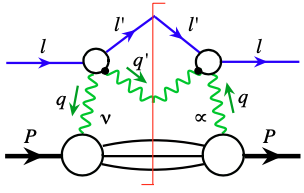} 
	\caption{The cut diagram representation of the QED radiative corrections to the inclusive DIS cross sections.
	}	
\label{fig:rc}
\end{figure}

However, the collision-induced photon radiation can change the momentum transfer between the colliding lepton and hadron from $q^\mu$ to $\hat{q}^\mu$, as shown in Fig.~\ref{fig:dis0}(b), making the characteristics of the hard probe experienced by the colliding hadron from (${x}_B,{Q}^2$) to ($\tilde{x}_B,\hat{Q}^2$) with 
\begin{equation}
\tilde{x}_B=\frac{\hat{Q}^2}{2P\cdot\hat{q}} \, ,
\quad \quad
\hat{Q}^2=-\hat{q}^2 \, ,
\label{eq:xbq2-parton}
\end{equation}
as discussed in details in Ref.~\cite{Liu:2021jfp}. 
Furthermore, with collision-induced radiation from the observed leptons, as shown in Fig.~\ref{fig:dis1}, the momentum of exchanged virtual photon $q^\mu$ is no longer fixed by the measured lepton momenta, $\ell$ and $\ell'$.  
%----------------------------------------------------------------
% Figure: NLO Inclusive DIS - BH
%----------------------------------------------------------------
%----------------------------------------------------------------
Instead, it becomes an integration variable whose value depends on available phase space of the radiated and unobserved photon of momentum $q'$.  From the square of the scattering amplitudes in Fig.~\ref{fig:dis1}, such non-trivial QED radiative correction to the DIS cross section can be represented by the cut-diagram in Fig.~\ref{fig:rc} 
%----------------------------------------------------------------
% Figure: RC to inclusive DIS
%----------------------------------------------------------------
%----------------------------------------------------------------
where the momentum of exchanged virtual photon $q^\mu$ could be perturbatively pinched to be on-shell at $q^2\sim 0$, 
\begin{eqnarray}
E'\frac{d\sigma^{\rm RC}_{e h\to e X}}{d^3\ell'} 
& \propto & 
\int d^4q \left[
L^{(1)}_{\mu\nu}(\ell,\ell',q) \frac{1}{q^2-i\epsilon} 
\right.
\nonumber \\
& & \hskip 0.5in \times \left.
\widetilde{W}_h^{\mu\nu}(P,q) \frac{1}{q^2+i\epsilon} 
\right]
\label{eq:qed_pinch} \\
& & \to \ \infty\, ,
\nonumber
\end{eqnarray}
when the scattered lepton momentum $\ell'_T$ is balanced by the unobserved photon of momentum $q'^\mu_T$ at this order.  
Having such an important contribution from QED, which is effectively an inclusive version of the Bethe-Heitler subprocess that competes with the Deep Virtual Compton Scattering process for extracting the generalized parton distributions (GPDs), should not be a surprise.
In order to keeping the exchanged photon to be virtual (as a hard probe for the colliding hadron) in the traditional RC approaches, parameter(s) have been introduced to limit the phase of the radiated and unobserved photon(s), which necessarily introduces uncertainty into the predictive power of the RC approaches.  

On the other hand, such pinched perturbative singularity is naturally taken care of in our joint QED and QCD factorization approach.  As demonstrated in this paper, for the region when $q^2\to 0$, the leading power contribution from the diagram in Fig.~\ref{fig:rc} to the DIS cross section in its $1/\ell'^2_T$ expansion can be approximated and factorized into a hard collision between the colliding lepton and an on-shell photon collinear to the colliding hadron, $e(\ell)+\gamma(q)\to e(\ell')+\gamma(q')$, convoluted with the photon distribution function of the hadron, while contribution from other region of $q^2$ should be perturbatively calculable as a part of NLO QED contribution to the factorized short-distance hard part of this joint factorization formalism.  In this joint factorization approach, we do not need to introduce any parameters other than the standard factorization scale.  Like other QCD factorization formalisms, predictive power of this joint factorization is from our ability to calculate the short-distance hard parts perturbatively in both QED and QCD and the universality of nonperturbative lepton and hadron distribution functions.

The proof of the joint QED and QCD factorization for inclusive DIS in lepton-hadron collisions, $e(\ell)+h(p)\to e(\ell')+X$, is effectively the same as the proof of QCD factorization for inclusive single hadron production in hadron-hadron collisions, $h_A(p_A)+h_B(p_B)\to h(p_h)+X$ at high transverse momentum $p_{h_T}$~\cite{Nayak:2005rt}. Since the photon, as a gauge boson of QED, does not carry charge and commutes with gluon fields of QCD, it is straightforward to verify that the same sequence of arguments used for justifying the QCD factorization of inclusive 
single high-$p_{h_T}$ hadron production in Ref.~\cite{Nayak:2005rt} is also valid for deriving the factorization formalism of inclusive single high-$\ell'_T$ lepton production in lepton-hadron collisions, which is the inclusive DIS when $\ell'_T \sim Q\gg\Lambda_{\rm QCD}$.  We have the corresponding factorization formalism~\cite{Liu:2020rvc,Liu:2021jfp},
\begin{eqnarray}
&&E'\frac{d\sigma^{\rm DIS}_{e(\ell) h(P)\to e(\ell') X}}{d^3\ell'} 
\approx \frac{1}{2S}\sum_{i,j,a}
\int_{\zeta_{\rm min}}^1 \frac{d\zeta}{\zeta^2}\, D_{e/j}(\zeta)
\nonumber\\
&& \hskip 0.5in 
\times
\int_{\xi_{\rm min}}^1 \frac{d\xi}{\xi}\, f_{i/e}(\xi)  \, 
\int_{x_{\rm min}}^1 \frac{dx}{x}\, f_{a/h}(x)
\label{eq:dis} \\
&& \hskip 0.5in
\times
(2\hat{s}) E_{k'}\frac{d\hat{\sigma}_{i(k) a(p)\to j(k') X}}{d^3 k'}\bigg|_{k=\xi\ell, p=xP, k'=\ell'/\zeta}
\nonumber \\
&& \hskip 0.3in 
\equiv  \frac{1}{2S}\sum_{i,j,a}
D_{e/j}(\zeta)\otimes_{\zeta} 
f_{i/e}(\xi) \otimes_{\xi}
f_{a/h}(x) 
\nonumber\\
&& \hskip 0.9in 
\otimes_x\,
\widehat{H}_{ia\to jX}(\hat{s},\hat{t},\hat{u}) \, ,
\label{eq:symbolic}
\end{eqnarray}  
where  $\otimes_i$ with $i=\zeta,\xi, x$ represent the convolution of momentum fraction of flavor ``$i$'' as defined in Eq.~(\ref{eq:dis}) and the short-distance partonic hard parts $\widehat{H}_{ia\to jX}$  in Eq.~(\ref{eq:symbolic}) are given by 
\begin{eqnarray}
&& \widehat{H}_{i(k)a(p)\to j(k')X}(\hat{s},\hat{t},\hat{u}) 
\nonumber \\
&&\hskip 0.2in  
\equiv (2\hat{s})
E_{k'}\frac{d\hat{\sigma}_{i(k)a(p)\to j(k')X}}{d^3k'}\bigg|_{k=\xi\ell, p=xP, k'=\ell'/\zeta}\, ,
\label{eq:hard}
\end{eqnarray}
which is proportional to partonic scattering cross section of $\sigma_{i(k)a(p)\to j(k')X}$ with all perturbative collinear divergences along the ``observed'' parton momenta, $k$, $p$ and $k'$, removed.  The  parton-level Mandelstam variables in Eqs.~(\ref{eq:symbolic}) and (\ref{eq:hard}) are given by, 
\begin{equation}
\begin{split}
\hat{s} &=(k+p)^2 = (x\xi) S \, , 
\\
\hat{t} &= (k-k')^2 = (\xi/\zeta) T \, ,
\\
\hat{u} &=(p-k')^2 = (x/\zeta) U \, .  
\end{split}
\label{eq:mandelstam-parton}
\end{equation}
where additional hadron-level Mandelstam variables are defined as,
\begin{equation}
\begin{split}
T & =(\ell - \ell')^2=-Q^2 \, ,
\\
U & =(P-\ell')^2= -(1-y)S \, .  
\end{split}
\label{eq:mandelstam-hadron}
\end{equation}
In Eq.~(\ref{eq:dis}), the indices $i,j,a$ run over all lepton and parton flavors, 
$D_{e/j}(\zeta)$ is a lepton fragmentation function (LFF) for a produced lepton (or a parton) of flavor $j$ to fragment to an observed lepton $e$ carrying $\zeta$ momentum fraction of the parent parton $j$, 
$f_{i/e}(\xi)$ is a lepton distribution function (LDF) for finding an active collinear lepton (or a parton) of flavor $i$ from the colliding lepton $e$ carrying its momentum fraction $\xi$, 
$f_{a/h}(x)$ is a parton distribution function (PDF) for finding an active collinear parton (or a lepton) of flavor $a$ and momentum $p=xP$ from the colliding hadron $h$ of momentum $P$, and
the dependence on the factorization scale ($\mu \sim \ell'_T$) is suppressed.
The joint QED and QCD factorization ensures that the partonic hard parts, 
 $\widehat{H}_{ia\to jX}(\hat{s},\hat{t},\hat{u})$ in Eq.~(\ref{eq:hard}), are infrared safe, and can be systematically and perturbatively calculated in a power series of $(\alpha_{em}^{m}\alpha_s^n)$ with $m\geq 2$ and $n\geq 0$.  
The integration limits in Eq.~(\ref{eq:dis}) are given by~\cite{Liu:2021jfp}
\begin{equation}
\begin{split}
\zeta_{\rm min} &= - \frac{T+U}{S}
= 1- (1-x_B)\, y \, , 
\\
\xi_{\rm min} &= - \frac{U}{\zeta\, S + T}
=\frac{1-y}{\zeta-x_B y}\, , 
\\
x_{\rm min} &= - \frac{\xi\, T}{\xi\, \zeta\, S + U}
=\frac{\xi \, x_B \, y}{\xi\zeta +y -1} \equiv \tilde{x}_B\, .
\end{split}
\label{eq:limits} 
\end{equation}
Corrections to the joint factorization formalism in Eq.~(\ref{eq:dis}) are suppressed by the inverse powers of the large momentum transfer to be ${\cal O}(1/\ell'^2_T)$.
 
By neglecting the mass of leptons and hadrons, we can express the momentum of the colliding lepton $\ell$, the colliding hadron $P$, and the scattered lepton $\ell'$ in terms of their light-cone components, defined as $v^\pm \equiv (v^0\pm v^z)/\sqrt{2}$ for a four-vector $v=(v^0,\vec{v}_T,v^z)$, in the c.m.~frame of the lepton-hadron collision,
\begin{equation}
\begin{split}
\ell_\mu &= (\sqrt{S/2},0^-,0_\perp)\, ,
\\
P_\mu &= (0^+,\sqrt{S/2},0_\perp)\, ,
\\
\ell'_\mu &= (\ell'_T\, e^{y_{\ell}}/\sqrt{2},
            \ell'_T\, e^{-y_{\ell}}/\sqrt{2}, \vec{\ell'}_T)\, ,
\end{split}
\label{eq:momenta}
\end{equation}
where $\ell'_T=\sqrt{\ell'^2_T}$ and $y_\ell=\ln(\ell'^+/\ell'^-)/2$ are the transverse momentum and the rapidity of the observed lepton, respectively.

Unlike in Eq.~(\ref{eq:dis0_sfs}), the joint factorization formula for inclusive DIS in Eq.~(\ref{eq:dis}) does not depend on the structure functions, which are the consequence of the one-photon exchange approximation.  All leading power non-perturbative physics in the $1/\ell'^2_T$ expansion are represented by the universal LDFs, LFFs and PDFs.  Like the PDFs, both LDFs and LFFs are in principle non-perturbative since a radiated photon can further split into a lepton pair as well as a pair of quark-antiquark at a non-perturbative scale.  The factorization scale dependence of LDFs and LFFs is determined by the DGLAP-type evolution equations with evolution kernels perturbatively calculated in both QED and QCD, which will be discussed further in Sec.~\ref{sec:numerical}.

The most important lowest order hard part in Eq.~(\ref{eq:dis}) is given by subprocess $e(k)+q(p)\to e(k')+q$~\cite{Liu:2021jfp}
\begin{eqnarray}
 \widehat{H}^{(2,0)}_{eq\to eX}
 &=&
 \left(2\hat{s}\right) \left[ E_{k'} \frac{d\sigma^{\rm (LO)}_{e q\to e q}}{d^3k'}  \right]
 \nonumber \\
 &=&
 e_q^2 \left(4\alpha_{em}^2\right)  \left[  \frac{\hat{s}^2 + \hat{u}^2 }{ \hat{t}^2 } \right] 
 \delta\left( \hat{s} + \hat{t}+\hat{u} \right)
 \label{eq:H20}  \\
&=&
 e_q^2 \left(4\alpha_{em}^2\right) \frac{x^2\zeta\left[(\zeta\xi S)^2+U^2\right]}{(\xi T)^2 \, (\zeta\xi S+U)}\,  \delta(x-\tilde{x}_B)\, ,
 \nonumber
\end{eqnarray}
where $e_q$ is the fraction of electric charge carried by a quark of flavor $q$.  

The next most relevant lowest-order hard part in Eq.~(\ref{eq:dis}) is $\widehat{H}^{(2,0)}_{\gamma q\to \gamma X}$ from subprocess, $\gamma(k)+q(p)\to \gamma(k')+q$,  when the center of mass energy of the collision of lepton-hadron is so high that the photon distribution of the colliding lepton and the photon fragmentation function of the observed lepton are sufficiently sizable.  We will explore the impact of these additional subprocesses in a future publication.

%================================================================
\section{Next-to-Leading Order Hard Parts from QCD}
\label{sec:qcd}
%================================================================

The NLO hard parts for the inclusive DIS cross section in Eq.~(\ref{eq:dis}) are given by QCD corrections at ${\cal O}(\alpha_{em}^2\alpha_s)$ and QED corrections at ${\cal O}(\alpha_{em}^3)$, which will be studied in the next section.  Due to the exchange of one photon at this order in QED, the QCD corrections are often presented in terms of short-distance contributions to the structure functions in Eq.~(\ref{eq:dis0_sfs}) which are factorized in terms of the hadron's PDFs~\cite{Roberts:1990ww,Tung:2001cv}.  

For calculating QCD corrections in this joint QED and QCD factorization approach while keeping short-distance QED corrections at the LO, we can simplify  the Eq.~(\ref{eq:dis}) as follows,
\begin{eqnarray}
E'\frac{d\sigma^{\rm DIS-QCD}_{e h\to e X}}{d^3\ell'} 
&=& \frac{1}{2S}
\int_{\zeta_{\rm min}}^1 \frac{d\zeta}{\zeta^2}\, D_{e/e}(\zeta)
\int_{\xi_{\rm min}}^1 \frac{d\xi}{\xi}\, f_{e/e}(\xi) 
\nonumber \\
&\ & {\hskip -0.9in} \times\sum_{a=q,\bar{q},g}
\int_{x_{\rm min}}^1 \frac{dx}{x}\, f_{a/h}(x) \,
\widehat{H}_{ea\to eX}(\hat{s},\hat{t},\hat{u}) \, ,
\label{eq:dis-qcd}
\end{eqnarray}  
which can be further reduced to $\sigma^{\rm Born}_{e h\to e X}$ in Eq.~(\ref{eq:dis0_sfs}) if we set $D_{e/e}(\zeta) = \delta(1-\zeta)$ and $f_{e/e}(\xi)=\delta(1-\xi)$.  With fully evolved and universal $D_{e/e}(\zeta)$ and $f_{e/e}(\xi)$, the difference between the $\sigma^{\rm DIS-QCD}_{eh\to eX}$ in Eq.~(\ref{eq:dis-qcd}) and $\sigma^{\rm Born}_{e h\to e X}$ in Eq.~(\ref{eq:dis0_sfs}) is the size of radiative corrections from the {\it process-independent} leading power collision-induced radiations in this joint QED and QCD factorization approach.  In the next section, we evaluate the {\it process-dependent}, but, perturbatively calculable contributions from the collision-induced QED radiations.

Taking the advantage of one-photon exchange and the expression for $\sigma^{\rm Born}_{e h\to e X}$ in Eq.~(\ref{eq:dis0_sfs}), the short-distance hard part of QCD contributions $\widehat{H}^{(2,n)}_{e(k)a(p)\to e(k')X}(\hat{s},\hat{t},\hat{u})$ in Eq.~(\ref{eq:dis-qcd}) can be expressed in terms of the available short-distance hard parts for the DIS structure functions, 
\begin{eqnarray}
&& \widehat{H}^{(2,n)}_{e(k)a(p)\to e(k')X}(\hat{s},\hat{t},\hat{u})  
=(2\hat{s})
 \frac{4\alpha_{em}^2}{\hat{y}(\hat{Q}^2)^2} 
 \label{eq:dis-qcd-sfs} \\
&&\hskip 0.5in \times
 \bigg[
\hat{x}_B\,\hat{y}^2 \hat{F}_{1/a}^{(n)}(\hat{x}_B,\hat{Q}^2) 
+ (1-\hat{y}) \hat{F}_{2/a}^{(n)}(\hat{x}_B,\hat{Q}^2)\bigg]
\nonumber
\end{eqnarray}
for unpolarized DIS.  In deriving Eq.~(\ref{eq:dis-qcd-sfs}), we neglected the term proportional to parton mass and have the kinematic variables defined as
\begin{equation}
\begin{split}
\hat{q}^{\mu} &= (k - k')^\mu = (\xi \ell - \ell'/\zeta)^\mu \, ,
 \\
\hat{Q}^2 &= -\hat{q}^2 = (\xi/\zeta)Q^2 = - \hat{t}\, ,
\\
\hat{x}_B &= \hat{Q}^2/2p\cdot \hat{q} = \hat{Q}^2/2 (xP)\cdot \hat{q} = \tilde{x}_B/x \, ,
\\
\hat{y} &= p\cdot\hat{q}/p\cdot k = \hat{Q}^2/(\hat{x}_B\hat{s}) = 1 + \frac{\hat{u}}{\hat{s}} \, .
\end{split}
\label{eq:dis-qcd-variables} 
\end{equation}
In Eq.~(\ref{eq:dis-qcd-sfs}), $\hat{F}_{i/a}^{(n)}(\hat{x}_B,\hat{Q}^2)$ with $i=1,2$ are partonic hard parts for the factorized hadron structure functions.  By substituting Eq.~(\ref{eq:dis-qcd-sfs}) into Eq.~(\ref{eq:dis-qcd}) and comparing it with Eq.~(\ref{eq:dis0_sfs}), we have 
\begin{align}
F_{1/h}(\tilde{x}_B,\hat{Q}^2)
&=\sum_{a,n=0} \int_{\tilde{x}_B}^1 \frac{dx}{x} f_{a/h}(x) \hat{F}_{1/a}^{(n)}(\frac{\tilde{x}_B}{x},\hat{Q}^2) 
\nonumber\\
F_{2/h}(\tilde{x}_B,\hat{Q}^2)
&=\sum_{a,n=0} \int_{\tilde{x}_B}^1 dx\, f_{a/h}(x) \hat{F}_{2/a}^{(n)}(\frac{\tilde{x}_B}{x},\hat{Q}^2) 
\label{eq:fac-sfs}
\end{align}
where $f_{a/h}(x)$ are PDFs with parton flavor $a=q,\bar{q}$, or $g$.  Since factorized short-distance hard parts do not depend on the details of the colliding hadron, the partonic hard parts can be perturbatively calculated by applying the factorization formula in Eq.~(\ref{eq:fac-sfs}) onto an asymptotic partonic state of flavor $a$ and momentum $p$, and expanding both sides of the formula order-by-order in the powers of coupling constant.  We derive QCD contributions to the short-distance hard parts at the power of $\alpha_s^n$ with $n\geq 1$,
\begin{eqnarray}
\hat{F}_{1/a}^{(n)}(\hat{x}_B,\hat{Q}^2)
&=& F_{1/a}^{(n)}(\hat{x}_B,\hat{Q}^2) 
\nonumber \\
& & \hskip -0.5in 
 - \sum_{m=1}^{n} \sum_{b} 
 \int_{\hat{x}_B}^1 \frac{dx}{x}
f_{b/a}^{(0,m)}(x)
\hat{F}_{1/b}^{(n-m)}(\frac{\hat{x}_B}{x},\hat{Q}^2) 
\nonumber\\
\hat{F}_{2/a}^{(n)}(\hat{x}_B,\hat{Q}^2)
&=& F_{2/a}^{(n)}(\hat{x}_B,\hat{Q}^2) 
\label{eq:qcd-hard-sfs} \\
& & \hskip -0.5in 
 - \sum_{m=1}^{n} \sum_{b} 
 \int_{\hat{x}_B}^1 dx\,
f_{b/a}^{(0,m)}(x)
\hat{F}_{2/b}^{(n-m)}(\frac{\hat{x}_B}{x},\hat{Q}^2) 
\nonumber
\end{eqnarray}
where $a,b=q,\bar{q},g$.  In Eq.~(\ref{eq:qcd-hard-sfs}), both the structure functions of a parton state $F^{(n)}_{i/a}$ with $i=1,2$ and the parton distribution functions of the same parton state $f_{b/a}^{(0,m)}$ have collinear divergences that are often regularized by dimensional regularization.  The QCD factorization ensures that all collinear divergences are cancelled between the first and the rest terms in the right-hand-side of Eq.~(\ref{eq:qcd-hard-sfs}), leaving $\hat{F}_{i/a}^{(n)}(\hat{x}_B,\hat{Q}^2)$ with $n\geq 1$ infrared safe order-by-order in QCD perturbation theory.  

\begin{figure}[htbp]
	\centering
	 \includegraphics[width=0.16\textwidth]{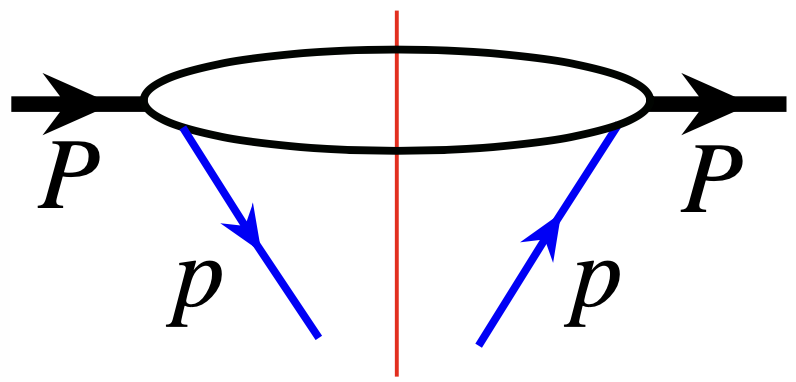} 
	 \hskip -0.2in $\longleftarrow \ \frac{\gamma^+}{2P^+}\delta\left( x-\frac{p^+}{P^+}\right)$
	\caption{The cut-diagram for quark distribution of momentum fraction $x$ along with corresponding cut-vertex. }	
\label{fig:cutdiagram-q}
\end{figure}

The subtraction terms in Eq.~(\ref{eq:qcd-hard-sfs}) depend on PDFs of flavor $b$ of an asymptotic parton state of flavor $a$, $f^{(0,n)}_{b/a}(x)$ with $a,b=q,\bar{q},g$, evaluated at ${\cal O}(\alpha_s^n)$ in QCD perturbation theory.   Since quarks carry both the color and electric charge, 
the quark distribution functions of a hadron of momentum $P$ in the joint QED and QCD factorization approach is required to be gauge invariant in both QED and QCD, and is slightly updated from its definition in QCD alone~\cite{Collins:1981uw}
\begin{eqnarray}
f_{q/h}(x,\mu^2) &=& 
\int \frac{d(P^+z^-)}{(2\pi)}\, e^{ixP^+z^-}
\label{eq:quark} 
\\
&  \times &
\langle h(P)|
\overline{\psi}_q(0) 
\frac{\gamma^+}{2P^+} \Phi^{\rm (F+U)}_{[0,z^-]} \psi_q(z^-)|h(P)\rangle
\nonumber
\end{eqnarray}
with $\mu$ representing both the renormalization and factorization scale and the gauge link,
\begin{equation}
\Phi^{\rm (F+U)}_{[0,z^-]}
= {\cal P}\, e^{
-i\int_0^{z^-} d\eta^-\left[
g_s A_g^{+a}(\eta^-)t^a - e A_{\gamma}^{+}(\eta^-)\right]
}
\label{eq:link-FU}
\end{equation}
where ${\cal P}$ denotes path ordering and $t^a$ is the generator for the fundamental representation of SU(3) color for QCD with color index $a=1,...8$ and quark color indices neglected, $g_s$ and $e$ are strong interaction charge and electric charge, respectively, the $A$ with subscripts, $g$ and $\gamma$, represent the gluon and photon fields, respectively, and superscripts, F and U, indicate the fundamental representation of SU(3) color for QCD and the U(1) gauge for QED, respectively. In momentum space, the same quark distribution function can be represented by the cut-diagram in Fig.~\ref{fig:cutdiagram-q} with the quark lines contracted with corresponding cut-vertex~\cite{Collins:1988wj}, where the gauge link was not explicitly shown.
%----------------------------------------------------------------
% Figure:  Cut diagram for quark distribution function
%----------------------------------------------------------------
%----------------------------------------------------------------

With the one-photon exchange for the QCD contributions, the structure functions of an asymptotic parton state $F^{(n)}_{i/a}$ in Eq.~(\ref{eq:qcd-hard-sfs}) can be extracted from the hadronic tensor $W^{\mu\nu}_a$ on the same partonic state.  From Eq.~(\ref{eq:sfs}), we obtain the projection operators for the unpolarized structure functions in $d=4-2\epsilon$ dimension,
\begin{eqnarray}
&& (1-\epsilon)F_{1/a}^{(n)}(\hat{x}_B,\hat{Q}^2)
\label{eq:sfs-f1n} \\
&& \hskip 0.4in =
\frac{1}{2}\left[-g^{\mu\nu}+\frac{4\hat{x}_B^2}{\hat{Q}^2} p^\mu p^\nu \right]
W_{a,\mu\nu}^{(n)}(\hat{q},p)
\nonumber \\
&& (1-\epsilon)F_{2/a}^{(n)}(\hat{x}_B,\hat{Q}^2)
\label{eq:sfs-f2n} \\
&& \hskip 0.4in =
\hat{x}_B\left[-g^{\mu\nu}+(3-2\epsilon)\frac{4\hat{x}_B^2}{\hat{Q}^2} p^\mu p^\nu \right]
W_{a,\mu\nu}^{(n)}(\hat{q},p)
\nonumber 
\end{eqnarray}
with $a=q,\bar{q},g$.

%----------------------------------------------------------------
% Figure:  Lowest order diagram contributing to the hadronic tensor
%----------------------------------------------------------------
\begin{figure}[htbp]
	\centering
	 \includegraphics[width=0.16\textwidth]{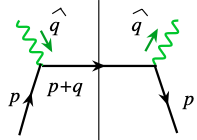} 
	\caption{The lowest order cut diagram contributing to the hadronic tensor of inclusive DIS.
	}	
\label{fig:dis-lo}
\end{figure}
%----------------------------------------------------------------
With the lowest order diagram contributing to the hadronic tensor in Fig.~\ref{fig:dis-lo}, we obtain,
\begin{eqnarray}
-g^{\mu\nu}\, W_{q,\mu\nu}^{(0)}(\hat{q},p) &=& e_q^2\,\delta(1-\hat{x}_B) \,
\nonumber\\
p^\mu p^\nu W_{q,\mu\nu}^{(0)}(\hat{q},p) &=&0 \, ,
\label{eq:dis-wmn0}
\end{eqnarray}
from which we have the short-distance contributions to the structure functions,
\begin{eqnarray}
\hat{F}_{1/q}^{(0)}(\hat{x}_B) &=& \frac{1}{2} e_q^2\, \delta(1-\hat{x}_B) \, ,
\nonumber \\
\hat{F}_{2/q}^{(0)}(\hat{x}_B) &=& \hat{x}_B\, e_q^2\, \delta(1-\hat{x}_B) \, ,
\label{eq:sts-f12-0} 
\end{eqnarray}
which confirms that the Callan-Gross relation~\cite{Callan:1969uq} $F_2(x_B)=2x_B F_1(x_B)$ is satisfied at this order.
From Eqs.~(\ref{eq:dis-qcd-sfs}) and (\ref{eq:sts-f12-0}), we obtain the lowest order hard part for the inclusive DIS,
\begin{eqnarray}
&& \widehat{H}^{(2,0)}_{eq\to eX}(\hat{s},\hat{t},\hat{u})  
\label{eq:dis-qcd-0} \\
&&\hskip 0.4in 
=e_q^2\, (4\alpha_{em}^2) \frac{1}{\hat{Q}^2}
\left[\frac{1+(1-\hat{y})^2}{\hat{y}^2}\right]
\delta(1-\hat{x}_B) \, ,
\nonumber
\end{eqnarray}
which can be easily verified to be the same as that in Eq.~(\ref{eq:H20}) by using the variables defined in Eq.~(\ref{eq:dis-qcd-variables}).

%----------------------------------------------------------------
% Figure: NLO Inclusive DIS - QCD
%----------------------------------------------------------------
\begin{figure}[htbp]
	\centering
	 \includegraphics[width=0.10\textwidth]{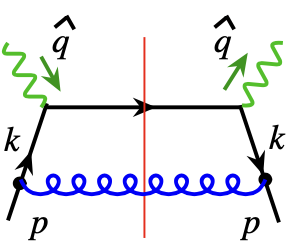} 
	 \hskip 0.1in
	 \includegraphics[width=0.10\textwidth]{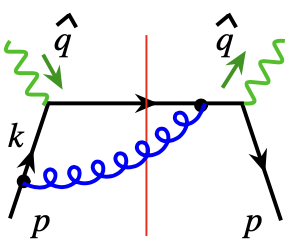} 
	 \hskip 0.1in
	 \includegraphics[width=0.10\textwidth]{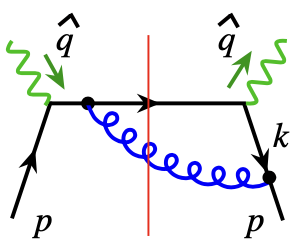} 
	 \hskip 0.1in
	 \includegraphics[width=0.10\textwidth]{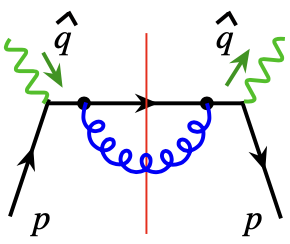} 
	 \\ 
	 (a)
	 \\
	 \includegraphics[width=0.165\textwidth]{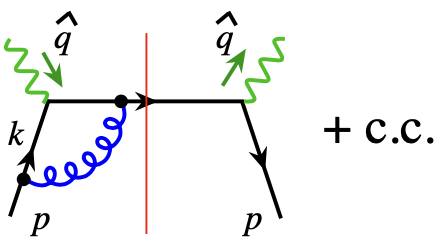} 
	 \\
	 (b)
\caption{Feynman diagrams for the NLO QCD contribution to inclusive DIS cross sections: (a) real, (b) virtual.
	}	
\label{fig:qcd-nlo}
\end{figure}
For NLO QCD contribution from an active quark, we evaluate both real and virtual contributions to the hadronic tensor $W^{(1)}_{q,\mu\nu}$ on the quark state of momentum $p=xP$ from diagrams in Fig.~\ref{fig:qcd-nlo}(a) and \ref{fig:qcd-nlo}(b), respectively in $d$-dimension.  After combing the real and virtual contributions and letting $d\to 4$ (or $\epsilon\to 0$), we obtain the following,
\begin{eqnarray}
&& -g^{\mu\nu}W^{(1)}_{q,\mu\nu}(\hat{q},p) 
= e_q^2(1-\epsilon) \left(\frac{\alpha_s}{2\pi}\right) 
\nonumber \\
&& \hskip 0.2in \times
\bigg\{
\bigg(-\frac{1}{\epsilon}\bigg) P^{(0,1)}_{q/q}(\hat{x}_B) \left(1+\epsilon\,\ln\left(4\pi e^{-\gamma_E}\right)\right)
\nonumber \\
&& \hskip 0.8in
+ P^{(0,1)}_{q/q}(\hat{x}_B) \ln\bigg[\frac{\hat{Q}^2}{\mu^2}\bigg]
\nonumber \\
&& \hskip 0.4in
+ C_F \bigg[
(1+\hat{x}_B^2)\bigg[\frac{\ln(1-\hat{x}_B)}{1-\hat{x}_B}\bigg]_+ -\frac{3}{2}\bigg[\frac{1}{1-\hat{x}_B}\bigg]_+ 
\nonumber \\
&& \hskip 0.8in 
- \frac{1+\hat{x}_B^2}{1-\hat{x}_B}\ln(\hat{x}_B) + 3 - \hat{x}_B 
\nonumber \\
&& \hskip 0.8in 
-\bigg[\frac{9}{2}+\frac{\pi^2}{3}\bigg]\delta(1-\hat{x}_B)\bigg]\bigg\} ,
\label{eq:qcd-nlo-g}
\end{eqnarray}
where color factor $C_F=(N_c^2-1)/(2N_c)$ with $N_c=3$, Euler constant $\gamma_E=0.5772...$ and the first order quark-to-quark splitting function in QCD is given by
\begin{equation}
P^{(0,1)}_{q/q}(x) = C_F\bigg[ \frac{1+x^2}{(1-x)_+} + \frac{3}{2}\delta(1-x)\bigg] \, ,
\label{eq:splitting_qq}
\end{equation}
with the ``+''-description defined as
\begin{align}
&\int_0^1 dx\frac{f(x)}{(1-x)_+} 
=\int_0^1 dx \frac{f(x) - f(1)}{1-x} \, ,
\label{eq:plus-description} \\
&\int_0^1 dx\, f(x)\left[\frac{\ln(1-x)}{(1-x)} \right]_+
\nonumber\\
& \hskip 0.6in
=\int_0^1 dx [f(x) - f(1)]\frac{\ln(1-x)}{1-x} \, ,
\label{eq:plus-log}
\end{align}
where $f(x)$ is a test function. 
With the same hadronic tensor $W^{(1)}_{q,\mu\nu}$, we also have 
\begin{eqnarray}
p^{\mu}p^{\nu}W^{(1)}_{q,\mu\nu}(\hat{q},p) &=& 
e_q^2 \, C_F \left(\frac{\alpha_s}{2\pi}\right) \frac{\hat{Q}^2}{4\hat{x}_B} \, .
\label{eq:qcd-nlo-pp}
\end{eqnarray}
From Eqs.~(\ref{eq:sfs-f1n}) and (\ref{eq:sfs-f2n}), we can derive NLO QCD contribution to the structure functions of a quark state, $F_{i/q}^{(1)}(\hat{x}_B.\hat{Q}^2)$ with $i=1,2$.  Their $(1/\epsilon)$ collinear divergence from Eq.~(\ref{eq:qcd-nlo-g}) should be canceled by the subtraction term in Eq.~(\ref{eq:qcd-hard-sfs}) as ensured by the factorization.  

%----------------------------------------------------------------
% Figure: cut-diagrams for the quark2quark pdf in QCD
%----------------------------------------------------------------
\begin{figure}[htbp]
	\centering
	 \includegraphics[width=0.12\textwidth]{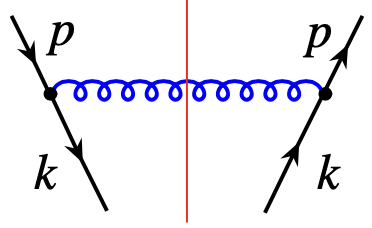} 
	 \hskip 0.25in
	 \includegraphics[width=0.12\textwidth]{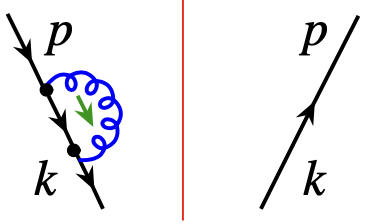}
	 \hskip 0.15in 
	 \includegraphics[width=0.12\textwidth]{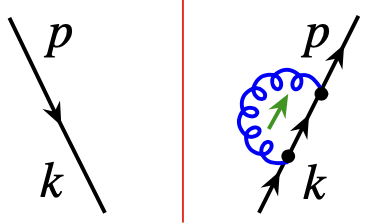}
	  \\
	  \caption{The cut-diagrams contribute to the ${\cal O}(\alpha_s)$ quark distribution of a quark state in QCD in a physical gauge.
	}	
\label{fig:q2q-qcd-1}
\end{figure}
%----------------------------------------------------------------
The subtraction term in Eq.~(\ref{eq:qcd-hard-sfs}) is proportional to $f_{q/q}^{(0,1)}$ - the ${\cal O}(\alpha_s)$ quark distribution of a quark state in QCD, which can be derived from its definition in Eq.~(\ref{eq:quark}), or by evaluating corresponding cut-diagrams in Fig.~\ref{fig:q2q-qcd-1} whose ultraviolet (UV) divergence is removed by a factorization scheme dependent UV-counter-term.  In the $\overline{\rm MS}$ factorization scheme, we have
\begin{eqnarray}
f_{q/q}^{(0,1)}(x) 
&=& 
\left(\frac{\alpha_s}{2\pi}\right) P^{(0,1)}_{q/q}(x) 
\nonumber \\
& & 
\times \bigg[
\bigg(-\frac{1}{\epsilon}\bigg) \left(1+\epsilon\,\ln\left(4\pi e^{-\gamma_E}\right)\right)
\bigg] .
\label{eq:fqq-qcd-1}
\end{eqnarray}
From Eq.~(\ref{eq:qcd-hard-sfs}) and Eqs.~(\ref{eq:sfs-f1n})-(\ref{eq:sfs-f2n}), we obtain NLO QCD contributions to the quark channel hard parts of factorized structure functions in the $\overline{\rm MS}$ scheme,
\begin{eqnarray}
&& \hat{F}_{1/q}^{(1)}(\hat{x}_B,\hat{Q}^2)
= \frac{1}{2}\, e_q^2 \left(\frac{\alpha_s}{2\pi}\right) 
\bigg\{
P^{(0,1)}_{q/q}(\hat{x}_B) \ln\bigg[\frac{\hat{Q}^2}{\mu^2}\bigg]
\nonumber \\
&& \hskip 0.3in
+ C_F \bigg[
(1+\hat{x}_B^2)\bigg[\frac{\ln(1-\hat{x}_B)}{1-\hat{x}_B}\bigg]_+ -\frac{3}{2}\bigg[\frac{1}{1-\hat{x}_B}\bigg]_+ 
\nonumber \\
&& \hskip 0.8in 
- \frac{1+\hat{x}_B^2}{1-\hat{x}_B}\ln(\hat{x}_B) + 3 
\nonumber \\
&& \hskip 0.8in
-\bigg[\frac{9}{2}+\frac{\pi^2}{3}\bigg]\delta(1-\hat{x}_B)\bigg]\bigg\} ,
\label{eq:qcd-nlo-f1} 
\end{eqnarray}
\begin{eqnarray}
&& \hat{F}_{2/q}^{(1)}(\hat{x}_B,\hat{Q}^2)
= \hat{x}_B\, e_q^2 \left(\frac{\alpha_s}{2\pi}\right) 
\bigg\{
P^{(0,1)}_{q/q}(\hat{x}_B) \ln\bigg[\frac{\hat{Q}^2}{\mu^2}\bigg]
\nonumber \\
&& \hskip 0.3in
+ C_F \bigg[
(1+\hat{x}_B^2)\bigg[\frac{\ln(1-\hat{x}_B)}{1-\hat{x}_B}\bigg]_+ -\frac{3}{2}\bigg[\frac{1}{1-\hat{x}_B}\bigg]_+ 
\nonumber \\
&& \hskip 0.8in 
- \frac{1+\hat{x}_B^2}{1-\hat{x}_B}\ln(\hat{x}_B) + 3 + 2\hat{x}_B
\nonumber \\
&& \hskip 0.8in
-\bigg[\frac{9}{2}+\frac{\pi^2}{3}\bigg]\delta(1-\hat{x}_B)\bigg]\bigg\} ,
\label{eq:qcd-nlo-f2} 
\end{eqnarray}
which indicates that the Callan-Gross relation is slightly violated for the short-distance contributions from the quark sector at this order. 
Substitute Eqs.~(\ref{eq:qcd-nlo-f1}) and (\ref{eq:qcd-nlo-f2}) into Eq.~(\ref{eq:dis-qcd-sfs}), we obtain NLO QCD contributions to partonic hard parts of inclusive DIS cross section from $eq\to eX$ subprocess in the joint QED and QCD factorization approach,
\begin{eqnarray}
&& \widehat{H}^{(2,1)}_{eq\to eX}(\hat{s},\hat{t},\hat{u})  
=e_q^2(4\alpha_{em}^2)\left(\frac{\alpha_s}{2\pi}\right)\frac{1}{\hat{Q}^2}
\nonumber \\
&& \hskip 0.15in \times \bigg\{
\frac{1+(1-\hat{y})^2}{\hat{y}^2}\bigg[ P^{(0,1)}_{q/q}(\hat{x}_B)\ln\bigg[\frac{\hat{Q}^2}{\mu^2}\bigg] 
\nonumber\\
&& \hskip 0.35in 
+ C_F\bigg(
(1+\hat{x}_B^2)\bigg[\frac{\ln(1-\hat{x}_B)}{1-\hat{x}_B}\bigg]_+ -\frac{3}{2}\bigg[\frac{1}{1-\hat{x}_B}\bigg]_+ 
\nonumber \\
&& \hskip 0.55in 
- \frac{1+\hat{x}_B^2}{1-\hat{x}_B}\ln(\hat{x}_B) + 3 
-\bigg[\frac{9}{2}+\frac{\pi^2}{3}\bigg]\delta(1-\hat{x}_B)\bigg)\bigg]
\nonumber\\
&& \hskip 0.3in 
+ \frac{1-\hat{y}}{\hat{y}^2}\bigg[ {C_F}\left(4\hat{x}_B\right)\bigg]\bigg\} ,
\label{eq:qcd-nlo-q}
\end{eqnarray}
which is infrared safe and perturbatively calculable without any free parameter other than the factorization scale $\mu^2$ in the $\overline{\rm MS}$ scheme.

In summary, the combination of Eqs.~(\ref{eq:dis-qcd-sfs}), (\ref{eq:qcd-hard-sfs}) and (\ref{eq:sfs-f1n})-(\ref{eq:sfs-f2n}) provides a systematic way to convert all available calculations of short-distance QCD contributions to the factorized DIS structure functions $\hat{F}_{i/a}^{(n)}(\hat{x}_B,\hat{Q}^2)$ with $i=1,2$ and $a=q,\bar{q},g$ at any order in powers of $\alpha_s^n$ to the corresponding perturbative hard parts $\widehat{H}^{(2,n)}_{ea\to eX}(\hat{s},\hat{t},\hat{u})$ in this joint QED and QCD factorization approach. 

%================================================================
\section{Next-to-Leading Order Hard Parts from QED}
\label{sec:qed}
%================================================================

In this section, we present the NLO QED contributiosn to the leading subprocess, $e(k)+q(p)\to e(k')+X$, of $E' d\sigma^{\rm DIS}_{e h\to e X}/d^3\ell'$ in Eq.~(\ref{eq:dis}).  The NLO QED contributions to other subprocesses and their numerical impact are expected to be less significant and will be presented in a future publication.

Unlike the QCD contributions, the approximation of one-photon exchange between colliding lepton and hadron is not valid when we calculate QED contributions beyond the LO, since photon can radiate from both lepton and quark which carries electromagnetic charge. Therefore, the concept of hadronic tensor and corresponding structure functions that we used to calculate the QCD contributions in the last section cannot be used for the calculations of NLO QED contributions. 

Like the QCD contributions, QED contributions to the partonic hard parts $\widehat{H}^{(m,0)}_{ia\to jX}$ in Eq.~(\ref{eq:dis}) are not sensitive to the details of colliding and scattered lepton and colliding hadron.  That is, the partonic hard parts $\widehat{H}^{(m,0)}_{ia\to jX}$ with $m\geq 2$ can be calculated perturbatively by applying the factorization formula in Eq.~(\ref{eq:dis}) onto asymptotic lepton and/or parton states: $i,j = e,\bar{e},\gamma$, other leptons or partons, and $a=q,\bar{q},g$, or leptons, and expanding both sides of the formula order-by-order in powers of coupling constants similar to the derivation of Eq.~(\ref{eq:qcd-hard-sfs}). 

To compute the NLO QED contributions to the hard part $\widehat{H}^{(3,0)}_{eq\to eX}$ in Eq.~(\ref{eq:dis}), we replace the colliding hadron $h$ in Eq.~(\ref{eq:dis}) by an asymptotic quark state of momentum $p=xP$, expand both sides of the equation to order $\alpha_{em}^3$, and obtain symbolically, 
\begin{equation}
\begin{split}
\sigma_{eq\to eX}^{(3,0)} &=
 D_{e/e}^{(0,0)}\otimes_\zeta f_{e/e}^{(0,0)}\otimes_\xi \widehat{H}^{(3,0)}_{eq\to eX}\otimes_x f_{q/q}^{(0,0)}
 \\
&+
D_{e/e}^{(1,0)}\otimes_\zeta f_{e/e}^{(0,0)}\otimes_\xi \widehat{H}^{(2,0)}_{eq\to eX}\otimes_x f_{q/q}^{(0,0)}
\\
&+
D_{e/e}^{(0,0)}\otimes_\zeta f_{e/e}^{(1,0)}\otimes_\xi \widehat{H}^{(2,0)}_{eq\to eX}\otimes_x f_{q/q}^{(0,0)}
\\
&+
D_{e/e}^{(0,0)}\otimes_\zeta f_{e/e}^{(0,0)}\otimes_\xi \widehat{H}^{(2,0)}_{eq\to eX}\otimes_x f_{q/q}^{(1,0)}
\\
&+
D_{e/e}^{(0,0)}\otimes_\zeta f_{e/e}^{(0,0)}\otimes_\xi \widehat{H}^{(2,0)}_{e\gamma\to eX}\otimes_x f_{\gamma/q}^{(1,0)} \, ,
\end{split}
\label{eq:qed1} 
\end{equation}
where $\sigma_{eq\to eX}^{(3,0)}\equiv (2\hat{s})E_{k'}d\sigma^{(3,0)}_{e(k)q(p)\to e(k')X}/d^3k'$ represents the partonic cross section at ${\cal O}(\alpha_{em}^{3})$ regularized in $d$-dimension.
With the zeroth order distribution and fragmentation functions, $D^{(0,0)}_{e/e}(\zeta)=\delta(1-\zeta)$, $f^{(0,0)}_{e/e}(\xi)=\delta(1-\xi)$, and $f^{(0,0)}_{q/q}(x)=\delta(1-x)$, we obtain the following expression for calculating the NLO QED contributions to the short-distance hard parts of the leading subprocess,
\begin{eqnarray}
\widehat{H}^{(3,0)}_{eq\to eX} &=&
\sigma_{eq\to eX}^{(3,0)} -
D^{(1,0)}_{e/e}\otimes_{\zeta} \widehat{H}^{(2,0)}_{eq\to eX} 
\nonumber\\
&-&
f^{(1,0)}_{e/e}\otimes_{\xi} \widehat{H}^{(2,0)}_{eq\to eX} 
- 
f^{(1,0)}_{q/q}\otimes_{x} \widehat{H}^{(2,0)}_{eq\to eX} 
\nonumber \\
&-& 
f^{(1,0)}_{\gamma/q}\otimes_{x} \widehat{H}^{(2,0)}_{e\gamma\to eX} \, ,
\label{eq:H30}
\end{eqnarray}
where the lowest order hard part $\widehat{H}^{(2,0)}_{eq\to eX}$ is given in Eq.~(\ref{eq:H20}) or (\ref{eq:dis-qcd-0}), 
%----------------------------------------------------------------
% Figure: Compton scattering on the colliding electron
%----------------------------------------------------------------
\begin{figure}[htbp]
	\centering
	 \includegraphics[width=0.12\textwidth]{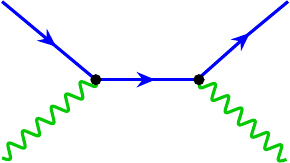} 
	 \hskip 0.5in
	  \includegraphics[width=0.12\textwidth]{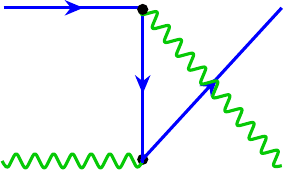} \\
	  \caption{Diagrams contribute to the LO hard part: $\widehat{H}^{(2,0)}_{e\gamma\to eX}$.
	}	
\label{fig:compton-e}
\end{figure}
%----------------------------------------------------------------
and $\widehat{H}^{(2,0)}_{e\gamma\to eX}$ is the LO contribution of the subprocess, $e(k)+\gamma(p) \to e(k')+X$.  Evaluating the LO diagrams in Fig.~\ref{fig:compton-e}, we have
\begin{eqnarray}
\widehat{H}^{(2,0)}_{e\gamma\to eX}
 &=& 
 \left(4\alpha_{em}^2\right)  \left[  \frac{\hat{s}^2 + \hat{u}^2 }{ \hat{s} (- \hat{u})} \right] 
 \delta\left( \hat{s} + \hat{t}+\hat{u} \right)
 \label{eq:H20g}  \\
&=& 
 \left(4\alpha_{em}^2\right) \frac{ (\zeta\xi S)^2+U^2}{(\xi S)(-U) \, (\zeta\xi S+U)}\,  \delta(x-\tilde{x}_{B})\, .
 \nonumber
 \end{eqnarray}
In the joint QED and QCD factorization approach, it is the last term, $f^{(1,0)}_{\gamma/q}\otimes_{x} \widehat{H}^{(2,0)}_{e\gamma\to eX}$ in Eq.~(\ref{eq:H30}) that naturally removes the perturbative pinch singularity of $\sigma_{eq\to eX}^{(3,0)}$ near $q^2=0$, coming from phase space integration of the exchanged photon of momentum $q$ in Fig.~\ref{fig:rc}.

In Eq.~(\ref{eq:H30}), the $f_{q/q}^{(1,0)}$ is the leading-order (LO) distribution of quarks in QED, and its calculation is the same as the calculation of $f_{q/q}^{(0,1)}$ in QCD with the gluon in Fig.~\ref{fig:q2q-qcd-1} replaced by a photon.  Similar to Eq.~(\ref{eq:fqq-qcd-1}), we have in the $\overline{\rm MS}$ scheme,
\begin{eqnarray}
f_{q/q}^{(1,0)}(x) 
&=&  
\left(\frac{\alpha_{em}}{2\pi}\right) P^{(1,0)}_{q/q}(x) 
\nonumber \\
& & \times
\bigg[
\bigg(-\frac{1}{\epsilon}\bigg) \left(1+\epsilon\,\ln\left(4\pi e^{-\gamma_E}\right)\right)
\bigg] ,
\label{eq:fqq-qed-1}
\end{eqnarray}
where the splitting function $P^{(1,0)}_{q/q}(x) = e_q^2\,P^{(0,1)}_{q/q}(x)/C_F$. 

Similarly, $f^{(1,0)}_{e/e}(\xi)$ in Eq.~(\ref{eq:H30}) is the LO electron distribution function of a physical electron.  Beyond the LO, electron distribution function of a physical electron, $f_{e/e}(\xi,\mu^2)$ has the same operator definition as that in Eq.~(\ref{eq:quark}) with the quark field replaced by the electron field and the hadron state $|h(P)\rangle$ replaced by an electron state $|e(\ell)\rangle$.  From the electromagnetic gauge invariance of the electron distribution, the gauge link in Eq.~(\ref{eq:link-FU}) is simplified to have only the photon field.  The first order electron distribution function of an electron, $f^{(1,0)}_{e/e}(\xi)$, can be derived by using the same cut diagrams in Fig.~\ref{fig:q2q-qcd-1} with the gluon replaced by a photon and quark lines replaced by electron lines, and we have 
\begin{equation}
f_{e/e}^{(1,0)}(\xi) 
=
f_{q/q}^{(1,0)}(\xi) / e_q^2 \, .
\label{eq:fee-qed-1}
\end{equation}
For calculations beyond the NLO~\cite{Bodwin:2017wrc}, we will need quark distribution function of an electron, $f_{q/e}(\xi,\mu^2)$, which should have the same operator definition in Eq.~(\ref{eq:quark}) with the hadron state $|h(P)\rangle$ replaced by an electron state $|e(\ell)\rangle$.

%----------------------------------------------------------------
% Figure:  Cut diagram for photon distribution function
%----------------------------------------------------------------
\begin{figure}[htbp]
	\centering
	 \includegraphics[width=0.16\textwidth]{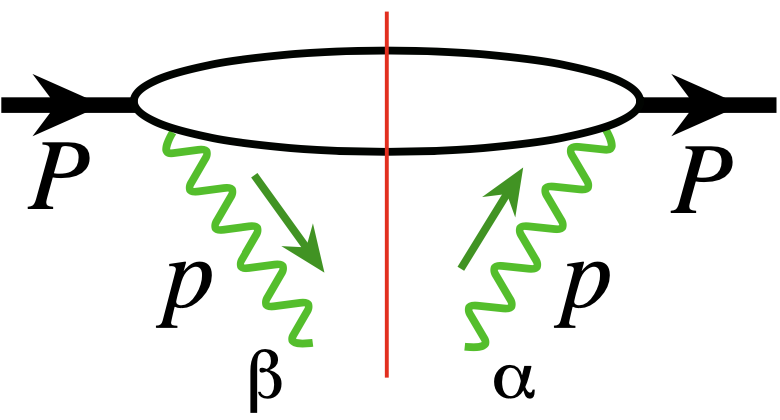} 
	 \hskip -0.2in $\longleftarrow \ {\cal P}_n^{\alpha\beta}(p)\, x\,\delta\left( x-\frac{p\cdot n}{P\cdot n}\right)$
	\caption{
    The cut-diagram for photon distribution of momentum fraction $x$ along with corresponding cut-vertex. }	
\label{fig:cutdiagram-gamma}
\end{figure}
%----------------------------------------------------------------
To evaluate the NLO QED contribution in Eq.~(\ref{eq:H30}), we also need a photon distribution function of an asymptotic quark state, $f^{(1,0)}_{\gamma/q}(x)$.  In general, the photon distribution function of a colliding hadron $h$ at a momentum fraction $x$, $f_{\gamma/h}(x,\mu^2)$, has a similar operator definition as the gluon distribution function of the hadron~\cite{Collins:1981uw} and can be represented by the cut-diagram in Fig.~\ref{fig:cutdiagram-gamma} in momentum space with corresponding cut-vertex for an off-shell photon of momentum $p$ where
\begin{equation}
{\cal P}_n^{\alpha\beta}(p)
= -g^{\alpha\beta}
+\frac{p^\alpha n^\beta + n^\alpha p^\beta}{p\cdot n}
- \frac{p^2}{(p\cdot n)^2} n^\alpha n^\beta
\label{eq:pol-tensor}
\end{equation}
with the light-cone vector $n^{\mu}=(n^+,n^-,n_\perp)=(0,1,0_\perp)$ and $p\cdot n = p^+$.  
Replacing the hadron by an asymptotic quark state, we derive the first order photon distribution function of a quark in the $\overline{\rm MS}$ scheme,
\begin{eqnarray}
f_{\gamma/q}^{(1,0)}(x) 
&=& 
\left(\frac{\alpha_{em}}{2\pi}\right) 
P^{(1,0)}_{\gamma/q}(x)
\nonumber \\
& & \times
\bigg[
\bigg(-\frac{1}{\epsilon}\bigg) \left(1+\epsilon\,\ln\left(4\pi e^{-\gamma_E}\right)\right)
\bigg] ,
\label{eq:fgammaq-qed-1}
\end{eqnarray}
with 
\begin{equation}
    P^{(1,0)}_{\gamma/q}(x) = e_q^2\, 
         \bigg[ \frac{1+(1-x)^2}{x} \bigg] .
         \label{eq:p_gaq}
\end{equation}
To calculate the factorized lepton-hadron DIS cross section in Eq.~(\ref{eq:dis}), we also need the photon distribution function of a physical electron, $f_{\gamma/e}(\xi,\mu^2)$, which has the same operator definition and cut-diagram representation as that of $f_{\gamma/h}(\xi,\mu^2)$ with the hadron state $|h(P)\rangle$ replaced by an electron state $|e(\ell)\rangle$.  Consequently, the first order photon distribution function of an electron is related to the first order photon distribution function of a quark as,
\begin{equation}
f_{\gamma/e}^{(1,0)}(\xi) 
= 
f_{\gamma/q}^{(1,0)}(\xi) / e_q^2 \, .
\label{eq:fgammae-qed-1}
\end{equation}

\begin{figure}[htbp]
	\centering
	 \includegraphics[width=0.11\textwidth]{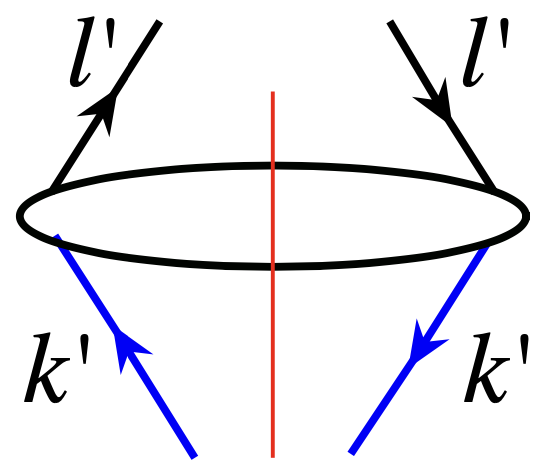} 
	 \hskip -0.1in $\longleftarrow \ \frac{\gamma^+}{4k'^+}\,\delta\left( \frac{1}{\zeta}-\frac{k'^+}{\ell'^+}\right)$
	\caption{The cut-diagram for lepton fragmentation function of momentum fraction $\zeta$ along with corresponding cut-vertex. }	
\label{fig:cutdiagram-lff-e}
\end{figure}

The $D_{e/e}^{(1,0)}$ in Eq.~(\ref{eq:H30}) is the LO electron fragmentation function of an electron.  In general, the LFF for a fermion (lepton or quark) of flavor $j$ to fragment into the observed electron $e$ of momentum $\ell'$, $D_{e/j}(\zeta,\mu^2)$, is similarly defined as the quark fragmentation function to an observed hadron~\cite{Collins:1981uw}
\begin{eqnarray}
D_{e/j}(\zeta,\mu^2) &=& \sum_{X} \int \frac{d(\ell'^+ z^-)}{2\pi}\, e^{-i\,\ell'^+ z^-/\zeta}
\nonumber \\
&& \hskip -0.4in \times 
{\rm Tr}\bigg[ \frac{\gamma^+}{4\,\ell'^+/\zeta}\langle 0| \overline{\psi}_j(0)\, \Phi^{\rm (F+U)}_{[0,\infty]} \, | e(\ell'),X\rangle
\label{eq:lff-e} \\
&& \hskip 0.2in \times
\langle e(\ell'),X |\, \psi_j(z^-)\, \Phi^{\rm (F+U)}_{[z^-,\infty]}\, | 0 \rangle \bigg]
\nonumber 
\end{eqnarray}
where the gauge link is defined in Eq.~(\ref{eq:link-FU}) for a quark to fragment to an electron, while for a lepton to fragment into an electron the gauge link  $\Phi^{\rm (F+U)}\to \Phi^{\rm (U)}$ with the U(1) gauge of QED. The LFF in Eq.~(\ref{eq:lff-e}) can also be represented by the cut-diagram in Fig.~\ref{fig:cutdiagram-lff-e} in momentum space with corresponding cut-vertex.
%----------------------------------------------------------------
% Figure:  Cut diagram for fermion fragmentation function
%----------------------------------------------------------------
%----------------------------------------------------------------
From ${\cal O}(\alpha_{em})$ contribution to the cut-diagram in Fig.~\ref{fig:cutdiagram-lff-e}, we derive the first order LFF 
$D_{e/e}^{(1,0)}(\zeta)$ in the $\overline{\rm MS}$ scheme, 
\begin{eqnarray}
D_{e/e}^{(1,0)}(\zeta) 
&=& 
\left(\frac{\alpha_{em}}{2\pi}\right) P^{(1,0)}_{e/e}(\zeta) 
\nonumber \\
& & \times
\bigg[
\bigg(-\frac{1}{\epsilon}\bigg) \left(1+\epsilon\,\ln\left(4\pi e^{-\gamma_E}\right)\right)
\bigg] ,
\label{eq:Dqq-qed-1}
\end{eqnarray}
where $P^{(1,0)}_{e/e}(\zeta) = P^{(1,0)}_{q/q}(\zeta)/e_q^2$. 

%----------------------------------------------------------------
% Figure: NLO real Feynman diagrams for QED contributions
%----------------------------------------------------------------
\begin{figure}[htbp]
	\centering
	     \begin{minipage}[t]{0.5\textwidth}
	          \centering
		 \includegraphics[width=0.8\textwidth]{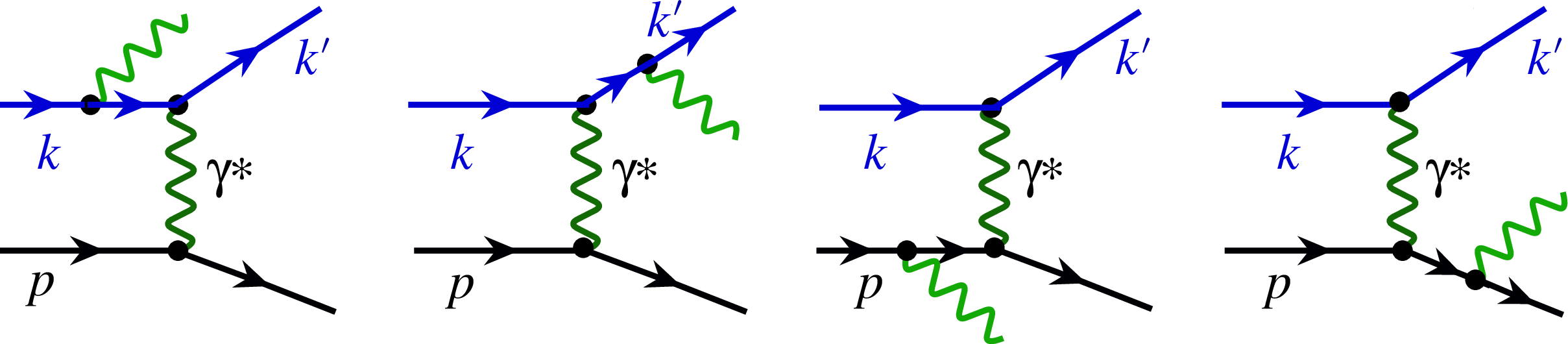} \\
		 (r1) \hskip 0.48in (r2) \hskip 0.48in (r3) \hskip 0.48in  (r4)
	     \end{minipage}
\caption{NLO Feynman diagrams for real QED contributions to $\sigma_{eq\to eX}^{(3,0)}$ in Eq.~(\ref{eq:H30}).
	              }
	\label{fig:qed_nlo_r}
\end{figure}
%----------------------------------------------------------------
\begin{figure}[htbp]
	\centering
	     \begin{minipage}[t]{0.5\textwidth}
	          \centering
		 \includegraphics[width=0.9\textwidth]{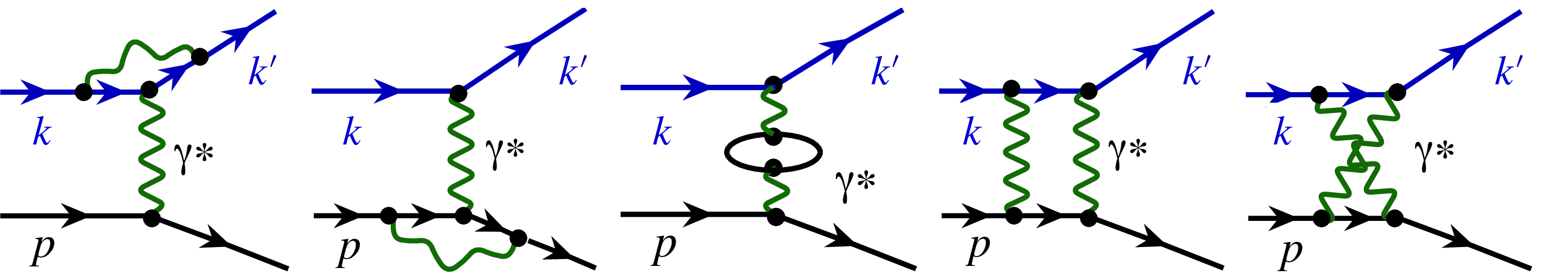} \\
		{\hskip -0.1in} (v1) \hskip 0.36in (v2) \hskip 0.36in (v3) \hskip 0.36in  (v4)  \hskip 0.36in  (v5)
	     \end{minipage}
\caption{NLO Feynman diagrams for virtual QED contributions to $\sigma_{eq\to eX}^{(3,0)}$ in Eq.~(\ref{eq:H30}).
	              }
	\label{fig:qed_nlo_v}
\end{figure}
%----------------------------------------------------------------

To evaluate the NLO QED contribution in Eq.~(\ref{eq:H30}), the last unknown is $\sigma_{eq\to eX}^{(3,0)}$, which is the NLO QED contributions to the leading subprocess, $e(k)+q(p)\to e(k')+q(p')$ of the inclusive lepton-hadron DIS.  Unlike the NLO QCD contributions to inclusive DIS, for which gluons can only be radiated from quarks, the NLO QED contributions can have photons radiated from both electron and quark, as shown in Fig.~\ref{fig:qed_nlo_r}, 
as well as have two-photon exchange between the electron and quark, as shown in Fig.~\ref{fig:qed_nlo_v}.  That is, the useful concept of hadronic tensor and structure functions for calculating QCD contributions to inclusive DIS as discussed in the last section, which was emerged naturally from the one-photon exchange, can not be applied for the calculations of factorized QED contributions to the inclusive DIS.  

Like the high order short-distance QCD contributions in hadronic collisions~\cite{Ellis:1979sj}, instead of the usual Mandelstam variables $(\hat{s},\hat{t},\hat{u})$, the $\sigma_{eq\to eX}^{(3,0)}$ is better calculated and represented with the partonic variables $(\hat{s},\hat{w},\hat{v})$ with the two dimensionless variables defined as,
\begin{equation}
\begin{split}
\hat{w} & \equiv  - \frac{\hat{u}}{\hat{s}+\hat{t}} = \frac{x\, WV}{\xi(x\, \zeta - 1 + V)} \, ,
\\
\hat{v} & \equiv  \frac{\hat{s}+\hat{t}}{\hat{s}} = \frac{x\, \zeta - 1 + V}{x\,\zeta}\, ,
\end{split}
\label{eq:def-wvhat}
\end{equation}
where corresponding dimensionless kinematic variables for observed lepton and hadron are given by,  
\begin{equation}
\begin{split}
W & \equiv  -\frac{U}{S+T} = \frac{1-y}{1-x_B\, y}\, ,   
\\
V & \equiv  \frac{S+T}{S} = 1-x_B\, y\, .
\end{split}
\label{eq:def-wv}
\end{equation}
The inclusive DIS cross section in Eq.~(\ref{eq:dis}) has an azimuthal symmetry with respect to the collision axis, and consequently, we should be able to represent the cross section by two independent variables, e.g., ($x_B,Q^2$), ($x_B,y$), ($y_{\ell}, \ell'^2_T$), or ($W,V$) defined above.  
In terms of the dimensionless variables ($W,V$), we have 
\begin{equation}
E' \frac{d\sigma^{\rm DIS}_{eh\to e' X}}{d^3\ell'} 
=
\frac{1}{\pi\, V S}\frac{d\sigma^{\rm DIS}_{eh\to e' X}}{dV dW}
\, .
\label{eq:dis_wv}
\end{equation}
Similar to Eq.~(\ref{eq:dis}), the factorization formalism can be expressed in terms of convolution of the parton-level dimensionless variables ($\hat{w},\hat{v}$),
\begin{eqnarray}
&&(1-V)WS\frac{d\sigma^{\rm DIS}_{eh\to e' X}}{dV dW}
\nonumber \\
&& \hskip 0.2in =
\sum_{i,j,a}
\int_{\zeta_{\rm min}}^1 d\zeta\, D_{e/j}(\zeta)
\int_{\hat{v}_{\rm min}}^{\hat{v}_{\rm max}} d\hat{v}\, f_{a/h}\left(\frac{1}{\zeta}\frac{1-V}{1-\hat{v}}\right) \,
\nonumber \\
&& \hskip 0.4in \times 
\int_{\hat{w}_{\rm min}}^1 d\hat{w}\, f_{i/e}\left(\frac{VW}{\zeta\hat{v}\hat{w}}\right) 
\hat{s}\frac{d\hat{\sigma}_{ia\to jX}}{d\hat{v} d\hat{w}} \, ,
\label{eq:fac_wv}
\end{eqnarray}  
where the integration limits are given by corresponding dimensionless variables,
\begin{equation}
\begin{split}
\zeta_{\rm min} 
&= 1-V + WV \, ,
\\
\hat{v}_{\rm min} 
&= \frac{WV}{\zeta} \, ,
\\
\hat{v}_{\rm max} 
&= 1 - \frac{1-V}{\zeta} \, ,
\\
\hat{w}_{\rm min}
&= \frac{WV}{\zeta\, \hat{v}} = \frac{\hat{v}_{\rm min}}{\hat{v}}\, .
\end{split}
\label{eq:wv_limits}
\end{equation}
The short-distance partonic cross section in Eq.~(\ref{eq:fac_wv}) can be expressed in terms of the perturbatively calculable hard part in Eq.~(\ref{eq:hard}),
\begin{equation}
\hat{s} \frac{d\hat{\sigma}_{i(k)a(p)\to j(k')X}}{d\hat{v} d\hat{w}}
=
\frac{\pi  \hat{v} \hat{s}}{2}\,
\widehat{H}_{i(k)a(p)\to j(k')X}(\hat{s},\hat{t},\hat{u}) \, .
\label{eq:hard_wv}
\end{equation}

In terms of these two dimensionless partonic variables $(\hat{w},\hat{v})$, the LO contribution to the partonic scattering process: $e(k)+q(p)\to e(k')+X$ 
is given by 
\begin{equation}
\hat{s} \frac{d\hat{\sigma}^{(2,0)}_{eq\to eq}}{d\hat{v} d\hat{w}}
= e_q^2 \left(4\alpha_{em}^2\right)\frac{\pi}{2}\left[\frac{1+\hat{v}^2}{(1-\hat{v})^2}\right] \delta(1-\hat{w})\, ,
\label{eq:lo_wv}
\end{equation}
and corresponding LO hard part,
\begin{eqnarray}
\widehat{H}_{eq\to eq}^{(2,0)}(\hat{s},\hat{t},\hat{u})
&=& \left(\frac{2}{\pi\hat{v}\hat{s}} \right) 
\hat{s} \frac{d\hat{\sigma}^{(2,0)}_{eq\to eq}}{d\hat{v} d\hat{w}}
\label{eq:H20_wv} \\
&=&
e_q^2 \left(4\alpha_{em}^2\right)\left[\frac{1+\hat{v}^2}{(1-\hat{v})^2}\right] \frac{1}{\hat{v}\hat{s}}\delta(1-\hat{w})\, .
\nonumber  
\end{eqnarray}
Using $\delta(1-\hat{w}) = (-\hat{u})\delta(\hat{s}+\hat{t}+\hat{u})$ and $\hat{v}=1+\hat{t}/\hat{s}$, it is easy to verify that $\widehat{H}_{eq\to eq}^{(2,0)}(\hat{s},\hat{t},\hat{u})$ in Eq.~(\ref{eq:H20_wv}) is the same as that in Eq.~(\ref{eq:H20}).

For deriving the NLO short-distance hard part $\widehat{H}^{(3,0)}_{eq\to eX}$ in Eq.~(\ref{eq:H30}), we evaluate contributions to 
\begin{equation}
\sigma^{(3,0)}_{eq\to eX} 
=  \left(\frac{2}{\pi\hat{v}\hat{s}} \right) 
\hat{s} \frac{d{\sigma}^{(3,0)}_{e(k)q(p)\to e(k')X}}{d\hat{v} d\hat{w}}
\label{eq:nlo_wv}
\end{equation}
by calculating perturbatively the real and virtual Feynman diagrams in Figs.~\ref{fig:qed_nlo_r} and \ref{fig:qed_nlo_v}, respectively.  Like the NLO partonic contributions to the structure functions in QCD, the $\sigma_{eq\to eX}^{(3,0)}$ has perturbative divergences.  We regularize all perturbative divergences by calculating these diagrams in $d$-dimension.  After combining the real and virtual contributions together, all perturbative infrared (IR) divergences are cancelled between the real and virtual diagrams, whose ultraviolet (UV) divergences are removed by the renormalization of QCD perturbation theory in the standard $\overline{\rm MS}$ UV-renormalization schedule.  The $\sigma_{eq\to eX}^{(3,0)}$ is left with perturbative collinear (CO) divergences from the phase space integration when the radiated photon is along the direction of observed partonic momenta, $k$, $k'$ and $p$, plus the situation when the exchanged photon is effectively on-shell and collinear to the colliding quark.  As ensured by the joint factorization, all residual perturbative CO divergences of $\sigma_{eq\to eX}^{(3,0)}$ are canceled by the subtraction terms in Eq.~(\ref{eq:H30}).  After adding together all terms on the right-hand-side of Eq.~(\ref{eq:H30}) and taking the dimension $d=4-2\epsilon \to 4$ (or $\epsilon \to 0$), we obtain the NLO short-distance contributions,
\begin{align}
&
\widehat{H}_{eq\to eq}^{(3,0)}(\hat{s},\hat{t},\hat{u})
=
e_q^2 \left(4\alpha_{em}^2\right) \frac{\alpha_{em}}{\pi} \,
\frac{1}{(1-\hat{v})^2\, \hat{s}}
\nonumber \\
& \hskip 0.1in \times
\bigg\{
e_l^2\, \dfrac{2(1+\vh^2)}{9\,\vh}
\bigg( 3\ln\bigg[\dfrac{(1-\vh)\hat{s}}{\mu^2}\bigg]-5\bigg)\delta(1-\wh)
\nonumber \\
& \hskip 0.2in
+ e_q 
\bigg(a_1\delta(1-\wh )+a_2 \dfrac{1}{[1-\wh]_+}
+a_4\bigg)
\nonumber \\
& \hskip 0.2in
+e_q^2 
\bigg({b_1} \delta (1-\wh)+{b_2} \dfrac{1}{[1-\wh]_+}+{b_3} \bigg[\dfrac{\ln (1-\wh)}{1-\wh}\bigg]_+
\nonumber \\
& \hskip 0.5in
+{b_4}\bigg)
\label{eq:nlo_qed} \\
& \hskip 0.2in
+{c_1} \delta (1-\wh)+{c_2} \dfrac{1}{[1-\wh]_+}
+{c_3} \bigg[\dfrac{\ln (1-\wh)}{1-\wh}\bigg]_+
+{c_4}
\bigg\}
\nonumber
\end{align}
where the perturbatively calculated coefficients $a_i$, $b_i$ and $c_i$ with $i=1,2,3,4$ are functions of dimensionless variables ($\hat{w},\hat{v}$), given in the Appendix A.  In Eq.~(\ref{eq:nlo_qed}), we organize the infra-safe hard contributions in terms of the power of quark fractional charge $e_q$, which is directly connected to the photon radiation from a quark.  The dependence on $e_q$ provides an extremely valuable information on how the collision-induced QED radiation contribute to the scattering cross sections, and how perturbative divergences are cancelled between the real and virtual diagrams and the interplay between the radiation from electrons and those from quarks.   

In Eq.~(\ref{eq:nlo_qed}), the term proportional to $e_l^2$ is from the one-loop virtual diagram in Fig.~\ref{fig:qed_nlo_v}(v3) interfered with the corresponding LO tree diagram of $eq\to eq$.  The $\mu$-dependence is from the UV renormalization of the one-loop photon vacuum polarization diagram in the $\overline{\rm MS}$ renormalization scheme.  The $e_l^2$ is given by, 
\begin{equation}
e_l^2 = \sum_f n_c^f \, e_f^2
\label{eq:el2}
\end{equation} 
where the sum of $f$ runs over all fermion flavors appeared in the photon vacuum polarization diagram in Fig.~\ref{fig:qed_nlo_v}(v3).  For a given flavor $f$, the unit of electric charge $e_f = 1$ for leptons and $e_f = e_q$ for quarks of fractional charge $e_q$.  Since there are three color states for a given quark flavor, the number of states for a given flavor in Eq.~(\ref{eq:el2}), $n_c^f = 1$ and $3$ for leptons and quarks, respectively. 
For our numerical calculation in the next section, we include two generation of fermions and neglect the charm quark mass in this vacuum polarization loop.

In Eq.~(\ref{eq:nlo_qed}), the term proportional to $e_q^3$ gets the virtual contributions from the so-called two-photon exchange diagrams in Figs.~\ref{fig:qed_nlo_v}(v4) and (v5) interfered with the corresponding LO tree diagram, and the real contributions from the interference between two real diagrams with the photon radiated from the electron in Figs.~\ref{fig:qed_nlo_r}(r1) and (r2) and those with the photon radiated from the quark in Figs.~\ref{fig:qed_nlo_r}(r3) and (r4).  Since the observed momentum of the scattered electron only fixes total momentum of the two exchanged photons in Figs.~\ref{fig:qed_nlo_v}(v4) and (v5), the virtual contribution from the two-photon exchange diagrams is IR sensitive when one of the two exchanged photon momenta becomes soft. However, it is the joint factorization formalism that ensures that such IR sensitivities are fully canceled by the real contributions from the interference between the real diagrams in Figs.~\ref{fig:qed_nlo_r}(r1) and (r2) and those in Figs.~\ref{fig:qed_nlo_r}(r3) and (r4), leaving the coefficient of $e_q^3$ completely IR-safe.  Since the one-loop two-photon exchange diagrams are UV finite, and the interference between the real diagrams with photon radiated from electron and quark does not generate CO divergence, the perturbatively calculated coefficients, $a_i$ with $i=1,2,3,4$ in Appendix A have no renormalization and factorization scale $\mu$-dependence. 

The contributions proportional to $e_q^4$ are from Feynman diagrams with photons radiated from quark lines including real diagrams of Figs.~\ref{fig:qed_nlo_r}(r3) and (r4), and virtual diagrams of Fig.~\ref{fig:qed_nlo_v}(v2), which are effectively the same diagrams as those for the NLO QCD contributions with the gluon replaced by the photon. That is, other than the color factor, this contribution is effectively the same as $\widehat{H}_{eq\to eq}^{(2,1)}(\hat{s},\hat{t},\hat{u})$ in Eq.~(\ref{eq:qcd-nlo-q}), which can be verified by recognizing that the one-photon exchange from the electron makes the dimensionless variables $\hat{v}$ and $\hat{w}$ not completely independent in the perturbatively calculated coefficients, $b_i$ with $i=1,2,3,4$ in Appendix A. 

The $e_q^2$ term in Eq.~(\ref{eq:nlo_qed}) corresponds to contributions from Feynman diagrams with the photon radiation only from the electron, like those in Figs.~\ref{fig:qed_nlo_r}(r1) and (r2), plus virtual diagrams in Figs.~\ref{fig:qed_nlo_v}(v1) and (v3).  This contribution should be closely related to the traditionally calculated RC factor, except it is completely IR-safe without the need to introduce any unknown parameter(s).  This IR-safety 
is ensured by the joint QED and QCD factorization.  As explained earlier, when the radiated photon is about back-to-back to the scattered electron, the contributions to the $e_q^2$ term have perturbative CO divergence from the phase space where exchange photon is about on-shell and collinear to the colliding quark.  Such perturbative CO divergence is avoided in the traditional RC approach by introducing a free parameter to limit the phase-space of the radiated photon(s), so that the collision-induced photon(s) is sufficiently collinear to the observed electron while keeping the exchanged photon (the hard probe) to be sufficiently virtual.  The choice of this unknown parameter is clearly sensitive to the detector capability to identify the collision-induced photon(s) at large angles, the specific final-state observed (and its kinematics) and the energy of collisions, as well as the order of perturbative calculations (or number of collision-induced photons).  On the other hand, such perturbative CO divergences are naturally removed order-by-order in our joint QED and QCD factorization formalism.  The coefficient of the $e_q^2$ term (all $c_i$ with $i=1,2,3,4$) in Eq.~(\ref{eq:nlo_qed}) is completely IR-safe without any free and unknown parameter, other than the standard factorization scale $\mu$.

The NLO factorized QED contribution to the leading subprocess $e(k)+q(p)\to e(k')+X$ of inclusive DIS in Eq.~(\ref{eq:nlo_qed}) provides an ideal example on how perturbative divergences can be systematically removed in the joint QED and QCD factorization approach.  The NLO factorized short-distance contributions to other subprocesses in Eq.~(\ref{eq:dis}) are expected to be smaller and can be evaluated systematically in the same way as what is presented here for evaluating $\widehat{H}_{eq\to eq}^{(3,0)}(\hat{s},\hat{t},\hat{u})$, and will be presented in the future publications.

%================================================================
\section{Numerical Results and Impacts}
\label{sec:numerical}
%================================================================

In this section, we study the numerical impact of the collision-induced QED and QCD radiation in the joint QED and QCD factorization approach.  We used CTEQ CT18 unpolarized PDFs~\cite{Hou:2019efy} for the following numerical calculations and did not see much difference from using other set of PDFs. 

Since the physical cross section should not depend on how do we factorize it perturbatively, the renormalization group improvement on the joint QED and QCD factorization formalism in Eq.~(\ref{eq:dis}) naturally leads to DGLAP-type evolution equations for PDFs, LDFs and LFFs.  Similar to the DGLAP evolution equation for PDFs in QCD~\cite{Dokshitzer:1977sg,Gribov:1972ri,Lipatov:1974qm,Altarelli:1977zs}
\begin{equation}
\mu^2 \frac{d}{d\mu^2} f_{i/h}(x,\mu^2) 
= \sum_{j} P_{i/j}\left(\frac{x}{x'},\alpha_s(\mu)\right) 
\otimes_{x'} f_{j/h}(x',\mu^2)
\label{eq:DGLAP}
\end{equation}
with $i,j=q,\bar{q},g$, DGLAP-type evolution equations for PDFs in the joint QED and QCD factorization approach have the same form as that in Eq.~(\ref{eq:DGLAP}), but, the parton indices $i,j$ need to be extended to include leptons, $l,\bar{l},\gamma$, with $l=e,\mu,\tau$ to include the evolution in both QCD and QED, as well as the mixing evolution from QCD to QED or vise versa~\cite{Liu:2021jfp,QW:evo,DeRujula:1979grv, Kripfganz:1988bd,Martin:2004dh,Manohar:2016nzj,Manohar:2017eqh}.  
The evolution kernels $P_{i/j}$ can be calculated perturbatively in both QED and QCD if the renormalization and factorization scale $\mu$ is sufficiently large,
\begin{equation}
    P_{i/j}(z,\mu^2) 
    =\sum_{m,n} \left[\frac{\alpha_{em}(\mu^2)}{2\pi}\right]^m\,
    \left[\frac{\alpha_s(\mu^2)}{2\pi}\right]^n P_{i/j}^{(m,n)}(z)\, ,
\label{eq:pij}
\end{equation}
where $m,n\geq 0$ with $m+n\geq 1$. Some of the LO kernels in Eq.~(\ref{eq:pij}) relevant to this work are given in the last two sections. Similar to the evolution equations of PDFs in Eq.~(\ref{eq:DGLAP}), we have DGLAP-type evolution equations for LDFs of an electron and LFFs to an electron, $f_{i/e}(\xi,\mu^2)$ and $D_{e/j}(\zeta,\mu^2)$, respectively, in this joint QED and QCD factorization approach.

Numerically, we found that the modification of PDFs from QED evolution is relatively smaller for the most regions of kinematic variables, which is consistent with what was observed earlier~\cite{Kripfganz:1988bd,Manohar:2016nzj,Manohar:2017eqh}, unless PDFs are modified significantly at the input scale nonperturbatively.  On the other hand, the impact of QCD evolution is more significant to the scale-dependence of LDFs and LFFs~\cite{QW:evo},
and in particular, LDFs and LFFs can no longer be evolved from the mass of electron to the hard scale like what was done for pure QED evolution~\cite{Hinderer:2015hra,Kniehl:1996we}. 
Like PDFs, LDFs and LFFs are non-perturbative in this joint QED and QCD factorization since QCD part of the evolution kernels $P_{i/j}$ are only valid at a hard and perturbative scale. 
While LDFs and LFFs are nonperturbative and their functional form at the input scale $\mu_0=m_c$ (the mass of Charm quark) should be extracted from experimental data, for the purpose of numerical calculations in this paper, we adopted, as a model or an approximation, that the input distributions for LDFs and LFFs at $\mu_0=m_c$ are chosen to be the same as those purbatively evolved from the mass of electron with purely QED evolution equations. We then use these QED evolved input distributions at $\mu_0$ to evolve the LDFs and LFFs to any factorization scale $\mu \geq \mu_0$ with the LO joint QED and QCD evolution kernels.
More detailed discussions will be presented in a future publication~\cite{QW:evo}.

To quantify the numerical impact of collision-induced radiations, as a reference, we define the DIS cross section {\it without} the collision-induced QED and QCD radiative contributions (LO in both QED and QCD hard coefficients), $d\sigma_{e(\ell)h(P)\to e(\ell') X}^{\rm LO-NR}$, by setting $D_{e/j}(\zeta)=\delta_{ej}\,\delta(1-\zeta)$ and $f_{i/e}(\xi)=\delta_{ie}\,\delta(1-\xi)$, and  $\widehat{H}_{ia\to j X} =\delta_{aq}\,\widehat{H}^{(2,0)}_{iq\to jX}$ in Eq.~(\ref{eq:dis}),
\begin{align}
E'\frac{d\sigma^{\rm LO-NR}_{e(\ell) h(P)\to e(\ell') X}}{d^3\ell'} 
& \nonumber \\
& \hskip -0.9in
\approx \frac{1}{2S}\sum_{q}
\int_{x_{\rm min}}^1 \frac{dx}{x}\, f_{q, /h}(x)
\widehat{H}^{(2,0)}_{eq\to eX}(\hat{s},\hat{t},\hat{u})
\label{eq:lo-nr}
\end{align}
with $\sum_q$ covers only accessible quark flavors of QCD.  
While keeping both QED and QCD hard coefficients at LO, we define the DIS cross section with only leading {\it process-independent} collision-induced QED radiative contributions by setting $i=e$, $j=e$, and $\widehat{H}_{ia\to j X} =\delta_{aq}\,\widehat{H}^{(2,0)}_{iq\to jX}$ in Eq.~(\ref{eq:dis}),
\begin{align}
E'\frac{d\sigma^{\rm LO-RC}_{e(\ell) h(P)\to e(\ell') X}}{d^3\ell'} 
&
\nonumber \\
& \hskip -0.9in
\approx \frac{1}{2S}
\int_{\zeta_{\rm min}}^1 \frac{d\zeta}{\zeta^2}\, D_{e/e}(\zeta)
\int_{\xi_{\rm min}}^1 \frac{d\xi}{\xi}\, f_{e/e}(\xi)  
\nonumber \\
& \hskip -0.8in
\times \sum_{q}
\int_{x_{\rm min}}^1 \frac{dx}{x}\, f_{q/h}(x)
\widehat{H}^{(2,0)}_{eq\to eX}(\hat{s},\hat{t},\hat{u})
\label{eq:lo-rc} 
\end{align}
where $f_{e/e}(\xi,\mu)$ and $D_{e/e}(\zeta,\mu)$ are perturbatively evolved and universal electron LDF and LFF, respectively.  While both LDFs and LFFs are universal but non-perturbative distribution functions that need to be extracted from experimental data, as a model, we could estimate the LDFs and LFFs by evaluating them purely in QED perturbatively.  Instead of calculating these functions in $d$-dimension as we did in last section, we can use electron mass as a natural regulator (or a cutoff) for the perturbative QED collinear singularity of LDFs and LFFs and calculate them in QED at $d=4$~\cite{Liu:2021jfp}
\begin{align}
f^{({\rm NLO},0)}_{e/e}(\xi,\mu^2) &=
 \delta(1-\xi) 
 \nonumber\\
 &+ \frac{ \alpha_{em}}{2\pi} \left[\frac{1+\xi^2}{1-\xi}\ln\frac{\mu^2}{(1-\zeta)^2m_e^2}\right]_+  ,
\label{eq:ldf} \\
D^{({\rm NLO},0)}_{e/e}(\zeta,\mu^2) &=
 \delta(1-\zeta) 
 \nonumber\\
 &+ \frac{ \alpha_{em}}{2\pi} \left[\frac{1+\zeta^2}{1-\zeta}\ln\frac{\zeta^2\mu^2}{(1-\zeta)^2m_e^2}\right]_+  ,
\label{eq:lff} 
\end{align}
where $m_e$ is electron mass. The perturbatively calculated LDF and LFF are only well-defined under the integration due to the ``$+$''-description and the $\delta$-function, and satisfy $\int_0^1d\xi\, f^{({\rm NLO},0)}_{e/e}(\xi,\mu^2) = 1$ and $\int_0^1 d\zeta\, D^{({\rm NLO},0)}_{e/e}(\zeta,\mu^2) = 1$.   

\begin{figure}[htbp]
	\centering
	\begin{tabular}{cc}
		\includegraphics[scale=0.5]{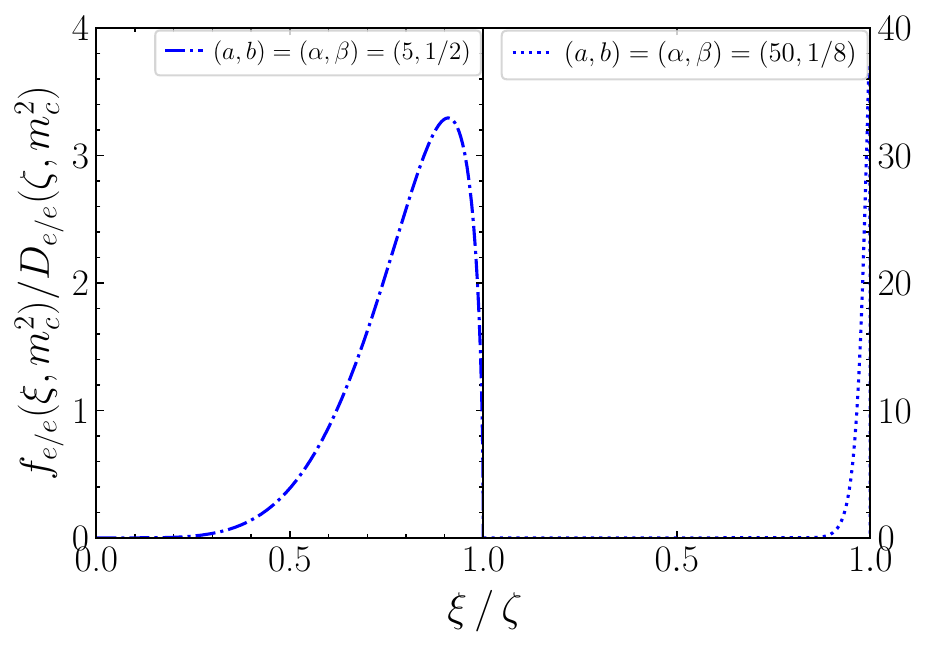}
	\end{tabular}
	\caption{Two very choices of input LDF and LFF at $\mu_0=m_c$. 
    }
	\label{fig:input-ldflff}
\end{figure}

Nonperturbatively, we expect $f_{e/e}(\xi,\mu^2) \to 0$ as $\xi\to 1$ and $D_{e/e}(\zeta,\mu^2) \to 0$ as $\zeta\to 1$, which is very different from their perturbative behavior in Eqs.~(\ref{eq:ldf}) and (\ref{eq:lff}), since the probability to find an electron carrying 100\% of its parent electron's momentum is vanishingly small once the electron is allowed to radiate photons.  Furthermore, unlike the probability to find a parton to carry the most momentum of parent hadron, which is small or unlikely, it is much more likely to find an electron to carry the most of parent electron's momentum.  That is, nonperturbative LDFs and LFFs at the input scale $\mu_0$ for the DGLAP-type evolution equations should more likely be peaked at a much larger momentum fraction, which is different from the typical behavior of light-flavor PDFs. To test the impact of collision-induced QED radiation to DIS cross sections due to the uncertainty of LDFs and LFFs, we parametrize the LDFs and LFFs at an input scale $\mu_0= m_c = 1.3$~GeV as 
\begin{align}
f_{e/e}(\xi,\mu_0^2) &=
\frac{\xi^a (1-\xi)^b}{B(1+a,1+b)}\, ,
\label{eq:input-ldf} \\
D_{e/e}(\zeta,\mu_0^2) &=
\frac{\zeta^\alpha (1-\zeta)^\beta}{B(1+\alpha,1+\beta)}\, .
\label{eq:input-lff}
\end{align}
With the Beta function, these parametrized LDFs and LFFs are normalized to have the same integrated probability as those in Eqs.~(\ref{eq:ldf}) and (\ref{eq:lff}).  While these non-perturbative and universal input distributions should be extracted by fitting experimental data, for the simplicity of the current discussion, we choose $f_{e/e}(\xi,\mu_0^2) = D_{e/e}(\zeta,\mu_0^2)$ with two sets of parameters: $(a,b)=(\alpha,\beta)=(50,1/8)$ or (5, 1/2), as shown in Fig.~\ref{fig:input-ldflff} to cover a wide range of LDFs and LFFs and their impact on the cross sections. For the other channels, we set $f_{\bar{e},\gamma/e}(\xi,\mu_0^2)=0$ and $D_{e/\bar{e},\gamma}(\zeta,\mu_0^2)=0$. 
%----------------------------------------------------------------
% Figure: Input LDF and LFF at $\mu_0$
%----------------------------------------------------------------
%----------------------------------------------------------------

\begin{figure}[htbp]
	\centering
	\begin{tabular}{cc}
		\includegraphics[scale=0.55]{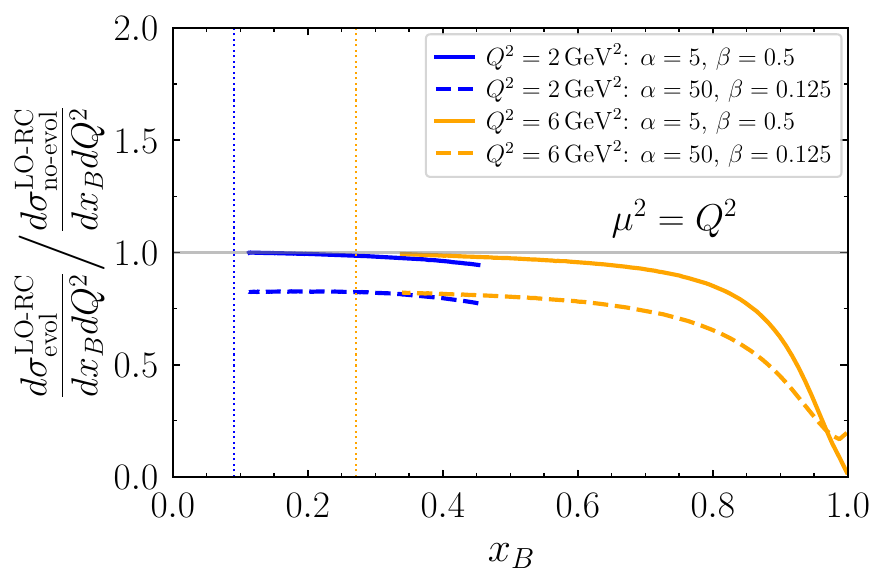} 
		\\ (a) \\
		\includegraphics[scale=0.55]{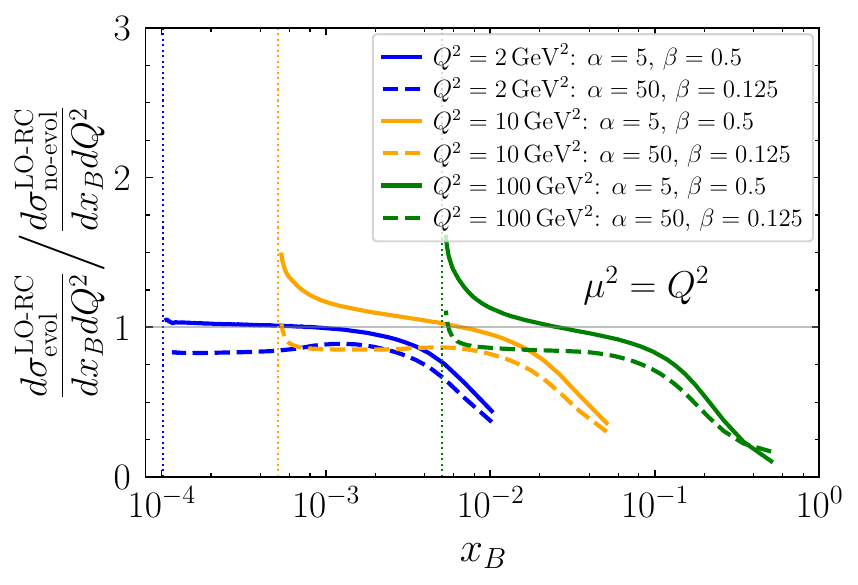} 
		\\ (b) 
	\end{tabular}
	\caption{$x_B$-dependence of the ratio of $\sigma_{eh\to eX}^{\rm LO-RC}$, defined in Eq.~(\ref{eq:lo-rc}), calculated with evolved LDFs and LFFs over that calculated with corresponding input LDFs and LFFs without evolution. 
    }
	\label{fig:r-dis-evolution}
\end{figure}

To test the impact of evolution, we plot the $x_B$-dependence of the ratio of DIS cross section $d\sigma_{eh\to eX}^{\rm LO-RC}$ in Eq.~(\ref{eq:lo-rc}), calculated with evolved over non-evolved LDFs and LFFs in Fig.~\ref{fig:r-dis-evolution} at $\sqrt{S}=4.7$~GeV (JLab energy) with $0.2\leq y \leq 0.8$ (a) and $\sqrt{S}=140$~GeV (EIC energy) with $0.01\leq y \leq 0.95$ (b), respectively.  The evolution was done with the leading-order evolution kernels and $\mu^2=Q^2$ was chosen for PDFs, LDFs and LFFs~\cite{QW:evo}. As expected, the effect of evolution is more significant for the steeper input distribution functions (dashed lines) even with a small amount of evolution. In addition, the numerical impact of evolution is enhanced at a small value of $x_B$, for which there is more phase space available for radiation, and in the larger $x_B$ region, where the distributions so as the cross sections are steeply decreasing.

%----------------------------------------------------------------
% Figure: Ratio of \sigma^{LO-RC} with evolved and non-evolved LDFs and LFFs
%----------------------------------------------------------------
%----------------------------------------------------------------

\begin{figure}[htbp]
	\centering
	\begin{tabular}{cc}
		\includegraphics[scale=0.55]{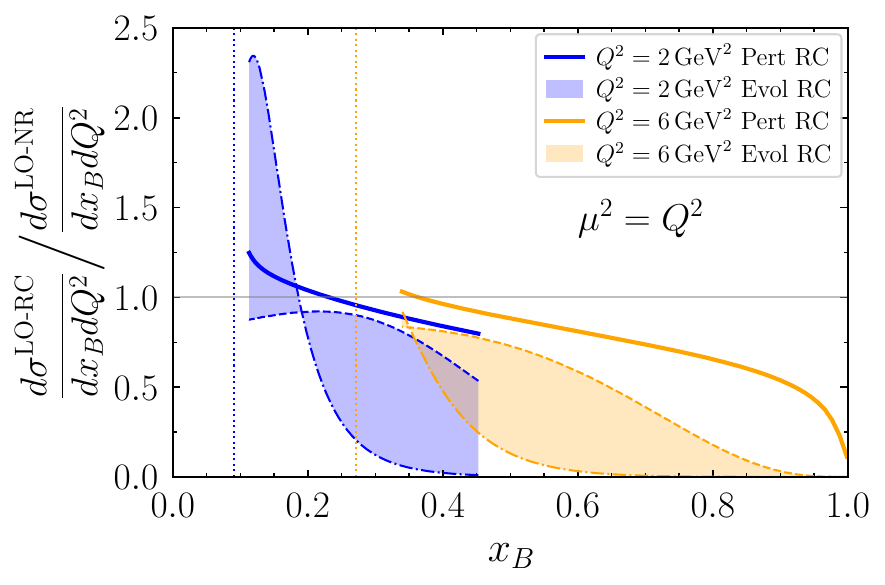} 
		\\ (a) \\
		\includegraphics[scale=0.55]{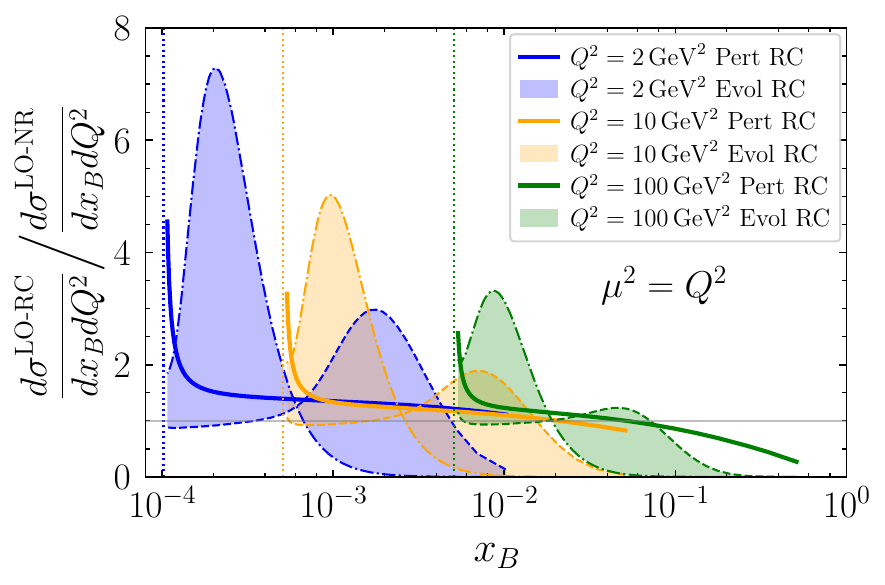} 
		\\ (b) 
	\end{tabular}
	\caption{$x_B$-dependence of the ratio of $\sigma_{eh\to eX}^{\rm LO-RC}$ over $\sigma_{eh\to eX}^{\rm LO-NR}$. 
    }
	\label{fig:r-dis-rad}
\end{figure}

To demonstrate the impact of {\it process-independent} collision-induced QED radiative contributions to the DIS cross sections, in Fig.~\ref{fig:r-dis-rad}, we plot the $x_B$-dependence of the ratio of DIS cross sections, $\sigma_{eh\to eX}^{\rm LO-RC}$ in Eq.~(\ref{eq:lo-rc}) over $\sigma_{eh\to eX}^{\rm LO-NR}$ in Eq.~(\ref{eq:lo-nr}), at $\sqrt{S}=4.7$~GeV (JLab energy) with $0.2\leq y \leq 0.8$ (a) and $\sqrt{S}=140$~GeV (EIC energy) with $0.01\leq y \leq 0.95$ (b), respectively.  
%----------------------------------------------------------------
% Figure: Ratio of \sigma^{LO-RC} over \sigma^{LO-NR} 
%----------------------------------------------------------------
%----------------------------------------------------------------
The shaded area covers the range of potential collision-induced QED contribution to the inclusive lepton-hadron DIS cross sections calculated with LDFs and LFFs evolved from two very different input LDFs and LFFs at $\mu_0=m_c$ with $(a,b)=(\alpha,\beta)=(50,1/8)$ or (5, 1/2), as shown in Fig.~\ref{fig:input-ldflff}.  The solid lines correspond to $\sigma_{eh\to eX}^{\rm LO-RC}$ calculated with the NLO LDF and LFF in Eqs.~(\ref{eq:ldf}) and (\ref{eq:lff}), respectively, which are effectively evolved in QED from a bare electron at $\mu_0=m_e$.  The range of the shadowed area and uncertainty in Fig.~\ref{fig:r-dis-rad} clearly shows that inclusive lepton-hadron DIS cross sections are very sensitive to collision-induced QED radiation and the details of the non-perturbative LDFs and LFFs. Like PDFs, it is the universality of LDFs and LFFs, once extracted from data of more physical observables~\cite{QW:evo}, that will allow us to make true predictions for physical observables in high energy collisions involving electron beam(s).

\begin{figure}[htbp]
	\centering
	\begin{tabular}{cc}
		\includegraphics[scale=0.55]{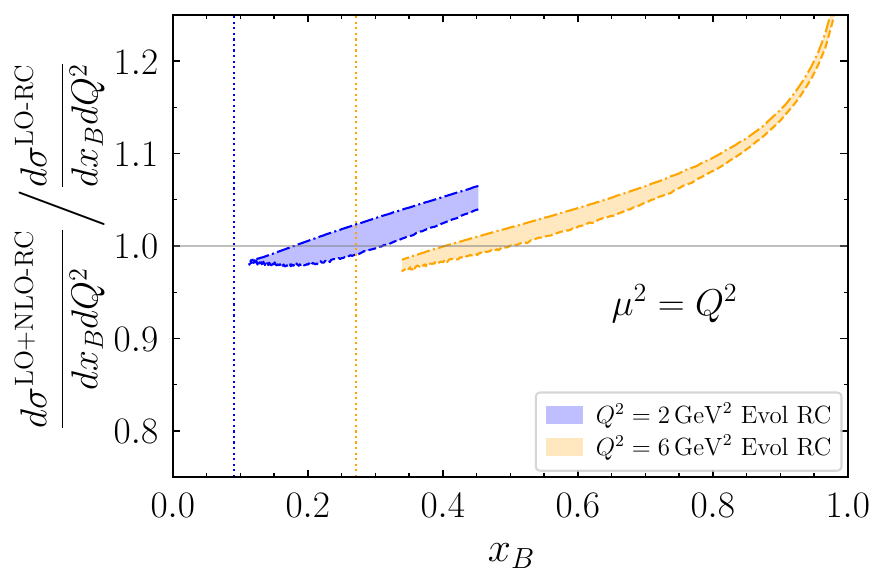} 
		\\ (a) \\
		\includegraphics[scale=0.55]{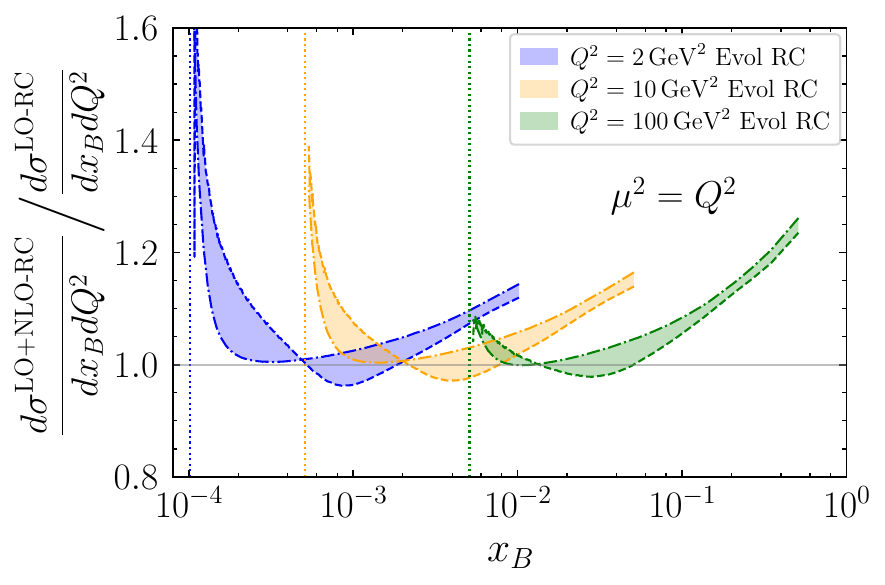} 
		\\ (b) 
	\end{tabular}
	\caption{$x_B$-dependence of the ratio of $\sigma_{eh\to eX}^{\rm LO+NLO-RC}$ over $\sigma_{eh\to eX}^{\rm LO-RC}$. 
    }
	\label{fig:r-nlo/lo-qed}
\end{figure}

To demonstrate the size of {\it process-dependent} collision-induced QED radiative contributions to the inclusive lepton-hadron DIS cross sections, we introduce,
\begin{align}
E'\frac{d\sigma^{\rm LO+NLO-RC}_{e(\ell) h(P)\to e(\ell') X}}{d^3\ell'} 
&
\nonumber \\
& \hskip -0.9in
\approx \frac{1}{2S}
\int_{\zeta_{\rm min}}^1 \frac{d\zeta}{\zeta^2}\, D_{e/e}(\zeta)
\int_{\xi_{\rm min}}^1 \frac{d\xi}{\xi}\, f_{e/e}(\xi)  
\nonumber \\
& \hskip -0.8in
\times \sum_{q}
\int_{x_{\rm min}}^1 \frac{dx}{x}\, f_{q/h}(x)
\widehat{H}^{\rm NLO}_{eq\to eX}(\hat{s},\hat{t},\hat{u})
\label{eq:nlo-rc} 
\end{align}
with $\widehat{H}^{\rm NLO}_{eq\to eX}(\hat{s},\hat{t},\hat{u})=\widehat{H}^{(2,0)}_{eq\to eX}(\hat{s},\hat{t},\hat{u}) + \widehat{H}^{(3,0)}_{eq\to eX}(\hat{s},\hat{t},\hat{u})$, keeping the QCD corrections to the hard part at the LO.  In Fig.~\ref{fig:r-nlo/lo-qed}, we plot the $x_B$ dependence of the ratio of DIS cross sections, $\sigma_{eh\to eX}^{\rm LO+NLO-RC}$ in Eq.~(\ref{eq:nlo-rc}) over $\sigma_{eh\to eX}^{\rm LO-RC}$ in Eq.~(\ref{eq:lo-rc}), at $\sqrt{S}=4.7$~GeV (JLab energy) with $0.2\leq y \leq 0.8$ (a) and $\sqrt{S}=140$~GeV (EIC energy) with $0.01\leq y \leq 0.95$ (b), respectively. For calculating both the numerator and the denominator, we used the same LDFs and LFFs evolved with LO evolution kernels.  Although $\alpha_{em}$ is a small number, the contributions of NLO collision-induced QED radiation to the short-distance perturbative hard part of this joint QED and QCD factorization, $\widehat{H}^{(3,0)}_{eq\to eX}(\hat{s},\hat{t},\hat{u})$, can be significant in influencing the DIS cross sections when there is more phase space for radiation in the small $x_B$ region or when the cross section is steeply falling at large $x_B$.
%----------------------------------------------------------------
% Figure: Ratio of \sigma^{LO+NLO-RC} over \sigma^{LO-RC} 
%----------------------------------------------------------------
%----------------------------------------------------------------

Unlike in Fig.~\ref{fig:r-dis-rad}, we cannot have the solid lines in Fig.~\ref{fig:r-nlo/lo-qed}, which are supposed to be derived by using perturbatively calculated LDF and LFF in Eqs.~(\ref{eq:ldf}) and (\ref{eq:lff}), respectively, to evaluate the DIS cross sections.  This is due to an intrinsic problem for evaluating a factorized expression with multiple perturbatively derived functions that have terms with the ``$+$''-description. 

Like all perturbatively calculated IR-safe distributions beyond the LO in QCD or QED, the perturbatively calculated NLO hard part in Eq.~(\ref{eq:H30}), LDF in Eqs.~(\ref{eq:ldf}) and LFF in Eq.~(\ref{eq:lff}) have terms proportional to $\delta$-function, terms with the ``$+$''-description, and possibly terms as analytical functions of various variables.  For evaluating the factorized expression in Eq.~(\ref{eq:fac_wv}), taking the term proportional to $[1/(1-\hat{w})]_+$ as an example, we need to evaluate the following integration,
\begin{align}
I\equiv 
&
\int_{\zeta_{\rm min}}^1 d\zeta\, D_{e/e}(\zeta)
\int_{\hat{v}_{\rm min}}^{\hat{v}_{\rm max}} d\hat{v}\, f_{q/h}\left(\frac{1}{\zeta}\frac{1-V}{1-\hat{v}}\right) \,
\nonumber \\
&\times
\int_{\hat{w}_{\rm min}}^1 d\hat{w}\, 
f_{e/e}\left(\frac{VW}{\zeta\hat{v}\hat{w}}\right) 
C(\hat{v},\hat{w})\frac{1}{[1-\hat{w}]_+}
\label{eq:divergence}
\end{align}  
where $C(\hat{v},\hat{w})$ is an analytical function of $\hat{v}$ and $\hat{w}$, made of the coefficients $a_i,b_i$ and $c_i$ in Eq.~(\ref{eq:H30}). The integration limits in Eq.~(\ref{eq:divergence}) are given in Eq.~(\ref{eq:wv_limits}). With the ``$+$''-description defined in Eq.~(\ref{eq:plus-description}), we can rewrite Eq.~(\ref{eq:divergence}) as
\begin{align}
I= &
\int_{\zeta_{\rm min}}^1 d\zeta\, D_{e/e}(\zeta)
\int_{\hat{v}_{\rm min}}^{\hat{v}_{\rm max}} d\hat{v}\, f_{q/h}\left(\frac{1}{\zeta}\frac{1-V}{1-\hat{v}}\right) \,
\nonumber \\
&\times\bigg\{
\int_{\hat{w}_{\rm min}}^1 d\hat{w}
\bigg[
f_{e/e}\left(\frac{\hat{w}_{\rm min}}{\hat{w}}\right) 
         \frac{C(\hat{v},\hat{w})}{1-\hat{w}}
\nonumber \\
& \hskip 0.8in
- f_{e/e}\left(\hat{w}_{\rm min}\right)  
         \frac{C(\hat{v},\hat{w}=1)}{1-\hat{w}}\bigg]
\label{eq:divergence00}\\
&  \hskip 0.4in
+
\ln(1-\hat{w}_{\rm min}) 
f_{e/e}\left(\hat{w}_{\rm min}\right) 
C(\hat{v},\hat{w}=1) \bigg\} 
\nonumber
\end{align}
with $\hat{w}_{\rm min}=
\hat{v}_{\rm min}/\hat{v}$ and $\hat{v}_{\rm min}=VW/\zeta$.
If we want to use the perturbatively calculated $f_{e/e}(\xi)$ in Eq.~(\ref{eq:ldf}) to evaluate Eq.~(\ref{eq:divergence00}), or more specifically, if $f_{e/e}(\hat{w}_{\rm min}) = \delta(1-\hat{w}_{\rm min})$, the first term in Eq.~(\ref{eq:ldf}), 
we would have 
\begin{align}
I\Rightarrow 
&
\int_{\zeta_{\rm min}}^1 d\zeta\, D_{e/e}(\zeta)
\int_{\hat{v}_{\rm min}}^{\hat{v}_{\rm max}} d\hat{v}\, f_{q/h}\left(\frac{1}{\zeta}\frac{1-V}{1-\hat{v}}\right)
\nonumber \\
& \hskip 0.1in \times
\ln\left(1-\frac{\hat{v}_{\rm min}}{\hat{v}}\right)  
\delta\left(1-\frac{\hat{v}_{\rm min}}{\hat{v}}\right)
C(\hat{v},\hat{w}=1)
\nonumber \\
& \rightarrow  \ \infty \, !
\label{eq:divergence11}
\end{align} 
That is, once we allow collision-induced radiation, which contribute to the hard parts perturbatively order-by-order, the LDFs and LFFs should vanish as corresponding momentum fraction goes to 1. The use of perturbative evaluated distribution functions, like those in Eqs.~(\ref{eq:ldf}) and (\ref{eq:lff}), together with perturbatively calculated hard parts beyond LO would lead to non-physical perturbative divergence. 

Corrections to the joint QED and QCD factorization formalism for inclusive and unpolarized lepton-hadron DIS in Eq.~(\ref{eq:dis}) is suppressed by the inverse power of the transverse momentum of the observed lepton, $1/\ell'^2_T$.  On the other hand, traditionally the Lorentz invariant $Q^2$ is the choice of hard scale for inclusive DIS. From Eq.~(\ref{eq:momenta}), we have
\begin{equation}
\ell'^2_T = (1-y)\, Q^2\, .
\label{eq:hard-scale}
\end{equation}
When the observed value $y = P\cdot (\ell-\ell')/P\cdot \ell = Q^2/(x_B S)$ small, the choice of $\ell'^2_T$ or $Q^2$ is effectively the same.  However, when the numerical value of $y$ is large, the value of $\ell'^2_T$ could become much smaller than the value of $Q^2$.  For example, for the lepton-hadron collider, such as the EIC, it is expected~\cite{Accardi:2012qut} to have $0.01 \leq y \leq 0.95$.  At $y=0.95$, $\ell'^2_T = Q^2/20$.  If $Q^2=1$~GeV$^2$ is considered a hard scale, as often referred to in the literature, the corresponding observed lepton transverse momentum, $\ell'_T$ can be as small as $220$~MeV.  This is not a hard scale!  

The value of $y$ plays a critical role in the relationship between the measured $Q^2$ and the true hard scale $\ell'^2_T$, as indicated Eq.~(\ref{eq:hard-scale}).  We included vertical dotted lines in Figs.~\ref{fig:r-dis-evolution}-\ref{fig:r-nlo/lo-qed} to indicate the value of $x_B$ corresponding to $y=1$ where $\ell'^2_T$ is effectively zero!

%================================================================
\section{Summary and Outlook}
\label{sec:conclusion}
%================================================================

The large momentum transfer between the colliding lepton and hadron in lepton-hadron scattering necessarily generates collision-induced QED radiation which could hinder our ability and precision to explore the internal structure of nucleons and nuclei, especially the confined 3-dimensional partonic structure of hadrons.  Instead of trying to improve the calculation of radiative correction factors to observables with different final-states and kinematics, we adapt a new joint QED and QCD factorization approach to evaluate the collision-induced QED radiation in the same way as we evaluate the collision-induced QCD radiation for all observables. Effectively, we treat the effect of collision-induced QED radiation, which depends on specific final-state and its kinematics, as a part of contributions to the production cross sections rather than a ``radiative correction" to the Born cross section without the QED radiation effect. 

In this paper, we presented the first calculation of NLO factorized QED and QCD contributions to the leading short-distance hard coefficients of inclusive lepton-hadron DIS in the joint QED and QCD factorization approach.  Unlike the traditional radiative correction approach to include QED radiation from leptons, we treat QED radiation from all charged particles, including quarks, equally and demonstrated that the NLO factorized QED contribution to the leading subprocess $e+q\to e+X$ is completely IR-safe and calculable without the need to introduce any parameters other than the standard factorization scale in the same way as the factorized QCD contribution.  Contributions to other subprocesses in this joint QED and QCD factorization approach are expected to be relatively small and will be presented in future publications.

With the perturbatively calculated short-distance hard parts and the universal LDFs, LFFs and PDFs, the lepton-hadron scattering with a large momentum transfer in this joint QED and QCD factorization approach does not need any free-parameter(s) other than the factorization scale.  Corrections to the factorization formalism in Eq.~(\ref{eq:dis}) are suppressed by the inverse powers of the large momentum transfer.  Instead of the traditional choice of $Q^2=-(\ell-\ell')^2$, we emphasize that it is critically important to have the transverse momentum of the observed lepton $\ell'_T$ sufficiently large; in particular, when $y$ is large, a larger $Q^2$ is needed to make $\ell'_T$ large enough, ensuring that the power corrections are sufficiently small.  

We emphasized that, like the PDFs, the universal LDFs and LFFs are nonperturbative in this joint QED and QCD factorization approach. The renormalization group improvement of the joint QED and QCD factorization formalism in Eq.~(\ref{eq:dis}) naturally led to the DGLAP-type evolution equations for PDFs as well as LDFs and LFFs.  Both QCD and QED contribute to the evolution of PDFs, LDFs and LFFs.  We found that the impact of QED evolution on PDFs is relatively small while the QCD evolution, due to its larger coupling constant, is more significant for the evolution of LDFs and LFFs, which should evolve from an input scale $\mu_0\sim m_c$ in this joint QED and QCD factorization approach, instead of $m_e$ as used in the pure QED evolution.  More detailed discussion on this joint evolution will be presented in a future publication~\cite{QW:evo}.  

We also demonstrated that this joint QED and QCD factorization is a natural approach to ensure that the contribution from the collision-induced QED radiation can be systematically evaluated without introducing any free parameter(s). As demonstrated in our calculation of $\widehat{H}_{eq\to eX}^{(3,0)}$, it is the photon distribution of the colliding hadron that helps to remove the perturbative pinch singularities of QED radiative contributions beyond the LO. 

Precision parity-violated DIS (PVDIS) provides opportunities to expand the phase space in the search of new interactions beyond the Standard Model, as well as to give a better access to partonic structure such as the isospin violation of quark distributions or strangeness content of the hadron~\cite{Ramsey-Musolf:2006evg, Londergan:1998ai, Londergan:2003ij, Sather:1991je}. Early experiments have clearly shown this capability of PVDIS~\cite{Prescott:1978tm, Prescott:1979dh, Souder:1990ia, SLACE158:2005uay}. The joint QED and QCD factorization approach to study the PVDIS will be able to help reduce the systematic theoretical uncertainty from collision-induced QED radiation, allowing us to have a better understanding of PVDIS so that we will have a precise and better way to probe the physics beyond the Standard Model. 

Although we introduced more universal unknown functions in this joint QED and QCD factorization approach to lepton-hadron scattering, such as LDFs and LFFs for describing inclusive lepton-hadron DIS, more factorizable observables in lepton-hadron scattering could provide additional information and constraints to help determine these universal functions~\cite{QW:evo}.  
Like QCD factorization, this joint QED and QCD factorization provides a robust approach to improve the precision of our calculations and our predictive power, allowing us to explore new observables, as well as the consistency in handling the collision kinematics.  Predictions come when we compare different factorizable observables depending on the same LDFs, LFFs and PDFs.

%================================================================
\section*{Acknowledgements}
%================================================================
We thank P.~Blunden, T.~Liu, W.~Melnitchouk, and N.~Sato
for helpful discussions and communications. 
This work is supported in part by the U.S. Department of Energy
(DOE) Contract No. DE-AC05-06OR23177, under which
Jefferson Science Associates, LLC operates Jefferson Lab.
K.W. is supported by JSPS KAKENHI Grant No. JP25K07286.

\newpage
%%%%%%%%%%%%%%%%%%%%%%%%%%%%%%%%%%%%%%%%%%%%%%%%%%%%%%%%
\appendix
%%%%%%%%%%%%%%%%%%%%%%%%%%%%%%%%%%%%%%%%%%%%%%%%%%%%%%%%

\begin{widetext}

%================================================================
\section{Hard coefficients of the next-to-leading order QED contributions}
\label{app:nlo-coefs}
%================================================================

The NLO short-distance contributions to the leading partonic subprocess, $e(k) + q(p) \to e(k') + X$, of inclusive lepton-hadron DIS cross sections, $\widehat{H}^{(3,0)}_{eq\to eX}$, is infrared safe and perturbatively calculable, and presented in Eq.~(\ref{eq:nlo_qed}) with corresponding coefficients $a_i$, $b_i$ and $c_i$ with $i=1,2,3,4$ given below,
 \begin{eqnarray} 
   a_1
   &=&\dfrac{\pi^2 \left(1-\vh^2\right)}{2 \vh}
   +\left(\dfrac{1}{\vh}-\vh\right) \ln ^2(1-\vh)-\left(\dfrac{9 \vh}{2}+\dfrac{7}{2 \vh}\right) \ln ^2\vh+\left(\dfrac{1}{\vh}-\vh\right) \ln (1-\vh)
   \nonumber\\
   &&
   +\left(5 \vh+\dfrac{3}{\vh}\right) \ln \vh \ln (1-\vh)-\left(\dfrac{1}{\vh}-1\right) \ln \vh \, ,
   \\
   a_2
   &=&-\dfrac{(1-\vh)^2}{1-\vh+\vh \wh}-\dfrac{2 (1-\vh)^2}{1-\vh \wh}+\dfrac{4 (1-\vh)^2}{(1-\vh \wh)^2}-\dfrac{2 (1-\vh)^2 \ln (1-\vh)}{1-\vh \wh}+\dfrac{2 (1-\vh) (2+\vh+3 \vh \wh) \ln \vh}{1-\vh \wh}
   \nonumber\\
   &&+\dfrac{2 (1-\vh) (1+\vh \wh) \ln (1-\wh)}{1-\vh \wh}+\dfrac{(1-\vh) (2+3 \vh+5 \vh \wh) \ln \wh}{1-\vh \wh}+(1-\vh) \ln (1-\vh+\vh \wh)
   \nonumber\\
   &&+2 (1-\vh) \ln (1-\vh \wh)-\dfrac{(1-\vh)^2}{\vh} \, ,
   \\
   a_3
   &=& 0 \, ,\\
   a_4
   &=&\dfrac{2 (1-\vh) \left(1-\vh^2\right) \ln (1-\vh)}{\vh (1-\vh \wh)}-\dfrac{\left(1-\vh^2\right) \ln (1-\vh+\vh \wh)}{\vh}-\dfrac{2 \left(1-\vh^2\right) \ln (1-\vh \wh)}{\vh}
   \nonumber\\
   &&-\dfrac{4 (1-\vh) \left(1+\vh^2\right) \ln \vh}{\vh (1-\vh \wh)}-\dfrac{(1-\vh) \left(3+5 \vh^2\right) \ln \wh}{\vh (1-\vh \wh)} \, ;
 \end{eqnarray}  
  
  \begin{eqnarray} 
   b_1
   &=&-\dfrac{\left(1+\vh^2\right)}{\vh} \left(\ln \dfrac{1-\vh}{\vh}-\dfrac{3}{4}\right) 
            \ln \left(\dfrac{\sh}{\mu ^2}\right)
          -\dfrac{\left(27+2 \uppi ^2\right) \left(1+\vh^2\right)}{12 \vh}-\dfrac{\left(1+\vh^2\right) \ln ^2(1-\vh)}{2 \vh} 
   \nonumber\\
   &&+\dfrac{\left(1+\vh^2\right) \ln ^2\vh}{2 \vh}+\dfrac{3 \left(1+\vh^2\right) \ln (1-\vh)}{2 \vh}-\dfrac{3 \left(1+\vh^2\right) \ln \vh}{4 \vh} \, ,
   \\
    b_2
    &=&\dfrac{(1-\vh)^3 \left(1+\vh^2\right)}{\vh (1-\vh \wh)^3} 
             \ln \left(\dfrac{\sh}{\mu ^2} \right)
           + \dfrac{(1-\vh)^3 \left(1+\vh^2\right)}{4 \vh (1-\vh \wh)^3} 
           \left(3 -4 \ln \vh \right) \, ,
    \\
    b_3
    &=&\dfrac{(1-\vh)^3 \left(1+\vh^2\right)}{\vh (1-\vh \wh)^3}   \, ,
    \\
    b_4
        &=&(1-\vh) \bigg(\dfrac{1}{2}-\dfrac{1+\vh}{2 (1-\vh \wh)}+\dfrac{1-\vh+\vh^2}{(1-\vh \wh)^2}+\dfrac{(1-\vh) \vh^2}{(1-\vh \wh)^3}\bigg) 
        \ln \left(\dfrac{ \sh }{\mu ^2}\right) 
    \nonumber\\
    && 
     + (1-\vh)\bigg[ \bigg(\dfrac{1}{2}-\dfrac{1+\vh}{2 (1-\vh \wh)}+\dfrac{1-\vh+\vh^2}{(1-\vh \wh)^2}+\dfrac{(1-\vh) \vh^2}{(1-\vh \wh)^3}\bigg) 
      \ln \left( (1-\wh) \vh\right) 
    \nonumber\\
    && \hskip 0.6in
   +\dfrac{3}{4}-\dfrac{3 (3-\vh)}{4 (1-\vh \wh)}+\dfrac{1+8 \vh-3 \vh^2}{4 (1-\vh \wh)^2}+\dfrac{(1-\vh) \left(5-3 \vh^2\right)}{4 (1-\vh \wh)^3}\bigg]\, ;
\end{eqnarray}
    
 \begin{eqnarray} 
   c_1
   &=& \dfrac{\left(1+\vh^2\right) (3+2 \ln \vh) }{2 \vh} 
           \ln \left( \dfrac{\sh}{\mu ^2} \right)
          + \dfrac{\left(1+\vh^2\right) (3+2 \ln \vh) }{2 \vh} \ln(1-\vh)
          -\dfrac{4 \left(1+\vh^2\right)}{\vh}+\dfrac{\left(1+\vh^2\right) \ln^2\vh}{2 \vh} \, ,
          \\
   c_2
   &=&\dfrac{2 \left(1-\vh\right)\left(1+\vh^2\right)}{\vh \wh (1-\vh \wh) (1-\vh+\vh \wh)} 
           \ln \left(\dfrac{\sh}{\mu ^2}\right)
          + \dfrac{2 \left(1-\vh\right)\left(1+\vh^2\right)}{\vh \wh (1-\vh \wh) (1-\vh+\vh \wh)} \ln(\vh)
          +\dfrac{2 \left(1-\vh\right)\left(1+\vh^2\right) \ln (1-\vh)}{\vh \wh (1-\vh \wh)} \, ,
   \\
   c_3
   &=&\dfrac{4 \left(1-\vh\right)\left(1+\vh^2\right)}{\vh \wh (1-\vh \wh)} \, ,
   \\
    c_4
    &=&
    \bigg[\dfrac{2-\vh+4 \vh^2+3 \vh^3}{2 (2-\vh) (1-\vh+\vh \wh)}+\dfrac{2-3 \vh+4 \vh^2+\vh^3}{(2-\vh) (1-\vh \wh)}
     -\dfrac{1+8 \vh-3 \vh^2+2 \vh^3}{2 \vh} -\dfrac{1+\vh^2}{\vh \wh}
     \nonumber\\
    && \hskip 0.4in
     -2 (1-\vh) \vh \wh-2 \vh^2 \wh^2 \bigg]
      \ln \left(\dfrac{\sh}{\mu ^2}\right)
    \nonumber\\
   &&  
      +  \bigg[\dfrac{2-\vh+4 \vh^2+3 \vh^3}{2 (2-\vh) (1-\vh+\vh \wh)}+\dfrac{2-3 \vh+4 \vh^2+\vh^3}{(2-\vh) (1-\vh \wh)}
     -\dfrac{1+8 \vh-3 \vh^2+2 \vh^3}{2 \vh} -\dfrac{1+\vh^2}{\vh \wh}
    \nonumber\\
    && \hskip 0.4in
     -2 (1-\vh) \vh \wh-2 \vh^2 \wh^2 \bigg]
      \ln \left(\vh \right)
    \nonumber\\
   && 
   +\dfrac{\left(1+\vh^2+2 \vh^2 \wh^2\right) \ln \wh}{\vh \wh}
   +\bigg[3-\dfrac{1+\vh^2}{\vh \wh}-2 \vh+\vh^2-2 \vh^2 \wh+2 \vh^2 \wh^2\bigg] \ln (1-\vh \wh)
   \nonumber\\
   &&  
   +\dfrac{(1-\vh)^2 (1-\wh) \left(1-\vh+\vh \wh+\vh^2 \wh\right)}{2 \vh \wh (1-\vh+\vh \wh)}+\dfrac{(1-\vh)^2 \ln (1-\vh+\vh \wh)}{1-\vh+\vh \wh}
   \nonumber \\
   && 
    -\bigg[5+\dfrac{(1-2\vh)(1+\vh^2)}{\vh \wh}-\dfrac{2\vh \left(1+\vh^2\right)}{1-\vh \wh}-2 \vh+\vh^2+2 (1-\vh) \vh \wh+2 \vh^2 \wh^2\bigg) 
    \ln (1-\vh)
    \nonumber \\
   &&  
   -\bigg[ \dfrac{1+12 \vh-3 \vh^2+2 \vh^3}{2 \vh}+\dfrac{(3-4\vh)(1+\vh^2)}{\vh \wh}
   -\dfrac{1+2 \vh+\vh^2+4 \vh^3}{1-\vh \wh}-\dfrac{(1-\vh)^2}{2 (1-\vh+\vh \wh)}
   \nonumber \\
   && \hskip 0.4in
   +2 (2-\vh) \vh \wh +2 \vh^2 \wh^2
   \bigg] \ln (1-\wh) \, .
\end{eqnarray}

\end{widetext}

%%%%%%%%%%%%%%%%%%%%%%%%%%%%%%%%%%%%%%%%%%%%%%%%
\bibliographystyle{elsarticle-num} 
\bibliography{reference}
%%%%%%%%%%%%%%%%%%%%%%%%%%%%%%%%%%%%%%%%%%%%%%%%

%%%%%%%%%%%%%%%%%%%%%%%%%%%%%%%%%%%%%%%%%%%%%%%%
\end{document}